\UseRawInputEncoding

\documentclass[journal]{IEEEtran}
\usepackage{cite}
\usepackage{graphicx,amssymb,lineno}
\usepackage{amsmath,amsfonts,amssymb}

\usepackage{algorithm}
\usepackage{algorithmic}
\usepackage[usenames]{color}
\usepackage{float}
\usepackage{mathtools}
\usepackage{cite}
\usepackage{booktabs}

\usepackage{graphicx,graphics,color,epsfig,subfigure,graphpap,rotate}
\usepackage{times, verbatim, subfigure, epsfig, graphicx, latexsym, amsmath}
\usepackage{url}
\usepackage{subfigure}

\begin{document}

\title{Resource Allocation for  Millimeter-Wave Train-Ground Communications in High-Speed Railway Scenarios}

\author{Xiangfei~Zhang,
        Yong~Niu,~\IEEEmembership{Member,~IEEE},
        Shiwen~Mao,~\IEEEmembership{Fellow,~IEEE},
        Yunlong~Cai,~\IEEEmembership{Senior Member,~IEEE},
        Ruisi~He,~\IEEEmembership{Senior Member,~IEEE},
        Bo~Ai,~\IEEEmembership{Senior Member,~IEEE},
        Zhangdui~Zhong,~\IEEEmembership{Senior Member,~IEEE},
        and Yiru~Liu

\thanks{Copyright (c) 2015 IEEE. Personal use of this material is permitted. However, permission to use this material for any other purposes must be obtained from the IEEE by sending a request to pubs-permissions@ieee.org. This study was supported by National Key R\&D Program of China (2020YFB1806903); in part by the National Natural Science Foundation of China Grants 61801016, 61725101, 61961130391, and U1834210; in part by the State Key Laboratory of Rail Traffic Control and Safety (Contract No. RCS2021ZT009), Beijing Jiaotong University; and supported by the open research fund of National Mobile Communications Research Laboratory, Southeast University (No. 2021D09); in part by the Fundamental Research Funds for the Central Universities, China, under grant number I20JB0200030; and supported by Frontiers Science Center for Smart High-speed Railway System; in part by the State Key Laboratory of Rail Traffic Control and Safety, Beijing Jiaotong University, under Grant RCS2019ZZ005; in part by the Fundamental Research Funds for the Central Universities 2020JBM089.}
\thanks{X. Zhang is with the State Key Laboratory of Rail Traffic Control and Safety, Frontiers Science Center for Smart High-speed Railway System, Beijing Jiaotong University, Beijing 100044, China, (e-mail: zxfei78@163.com).}
\thanks{Y. Niu is with the State Key Laboratory of Rail Traffic Control and Safety, Beijing Jiaotong University, Beijing 100044, China, and also with the National Mobile Communications Research Laboratory, Southeast University, Nanjing 211189, China (e-mail: niuy11@163.com).}
\thanks{S. Mao is with the Department of Electrical and Computer Engineering, Auburn University, Auburn, AL 36849-5201, USA (e-mail: smao@ieee.org).}
\thanks{Y. Cai is with the Department of ISEE, Zhejiang University, China (e-mail: ylcai@zju.edu.cn).}
\thanks{R. He, B. Ai, Z. Zhong, and Y. Liu are with the State Key Laboratory of Rail Traffic Control and Safety, Beijing Engineering Research Center of High-speed Railway Broadband Mobile Communications, and the School of Electronic and Information Engineering, Beijing Jiaotong University, Beijing 100044, China (e-mails: ruisi.he@bjtu.edu.cn; boai@bjtu.edu.cn; zhdzhong@bjtu.edu.cn; yrliu@bjtu.edu.cn).}
}

\maketitle

\begin{abstract}
With the development of wireless communication, higher requirements
arise
for train-ground wireless communications
in
high-speed railway (HSR) scenarios. The millimeter-wave (mm-wave) frequency band with rich spectrum resources can provide users in HSR scenarios with
high
performance
broadband multimedia services,
while
the full-duplex (FD) technology has become mature.
In this paper,
we study train-ground communication system performance in HSR scenarios with mobile relays (MRs) mounted on rooftop of train and operating in the FD mode.
We formulate a nonlinear programming problem to maximize network capacity by allocation of spectrum resources.
Then, we develop
a sequential quadratic programming (SQP) algorithm based on
the
Lagrange function
to solve the bandwidth allocation optimization problem
for
track-side base station (BS) and MRs in this mm-wave train-ground communication system.
Extensive simulation results demonstrate
that the
proposed
SQP algorithm can effectively
achieve high
network capacity
for
train-ground communication in HSR scenarios while
being robust to the residual self-interference (SI).
\end{abstract}

\begin{IEEEkeywords}
Full-duplex communications,
sequential quadratic programming,
Train-ground communications, high-speed railway, Millimeter-wave communications, Resource allocation.
\end{IEEEkeywords}

\section{Introduction}\label{S1}

Looking back at the past
decade,
the rapid development of high-speed railway (HSR) has
driven many technological innovations and changed people's lives.
As of the end of April 2015, the total operating mileage of the world's HSR reached 29,700 kilometers.
In particular,
as of the end of 2019, the operating mileage of Chinese HSR has reached 35,000 kilometers, ranking
the
first in the world.
Most of the Chinese HSR stations are located in the urban life circles
and the high-speed
trains have flexible schedules,
allowing
people
to travel
easily
to effectively save
costs and to improve the quality of people's lives.

On the other hand, since many passengers are accustomed to broadband wireless access in
their daily
living environment, more and more people hope to
have
high-quality broadband wireless access on mobile terminals.
However, in the HSR scenarios, the high-speed of train causes frequent handovers. When the cell radius is 1 to 2km, the train running at 350km/h will handover every 10 to 20s~\cite{Ai}. In addition, the rapid relative movement between the train and ground base station (BS) causes more serious Doppler shift and smaller channel coherence time. So the wireless channel in HSR scenarios has obvious non-stationary and fast time-varying characteristics, which seriously reduces the performance of train-ground communication systems~\cite{Doppler}. Moreover, due to the complexity and non-stationarity of HSR scenarios, there are weak field strength areas and blind areas, and the train body of metal material causes great penetration loss to the signal from the BS~\cite{HSRMR}.
Therefore,
meeting
passengers'
compelling
demand for broadband mobile communications
in the HSR environment has
become a key technical
challenge.
It
has become
particularly important to carry out research on broadband wireless communication technologies
in HSR scenarios. However, the currently widely used communication technology under HSR scenarios, such as GSM-R, can
only
support a rate that is too low to meet user's increasing demands for multimedia applications, and the current demand rate of each train is about 37.5Mbps. With the growth of
business
and entertainment
activities
and quality of service demands,
the demand rate
may
easily
reach 0.5--5Gbps in the
near
future~\cite{ROF}. Obviously, the current wireless transmission scheme
will be inadequate to
satisfy
the needs of HSR passengers.

To this end,
the 30--300GHz millimeter-wave (mm-wave) frequency band
can provide
rich spectrum resources for train-ground communications in HSR scenarios.
If mm-wave communications
are
used to carry broadband multimedia services, the transmission rate
in
the order of gigabit can be
offered~\cite{mmwave5G}. Mobile communication operators and wireless communication system designers are focusing on
developing the
related
technologies,
hoping to provide
users with
high-capacity broadband wireless access. Current typical application scenarios include high-speed data transmission between mobile devices, mm-wave wireless backhauls, military radar, autonomous cars, and
the co-design of radar and communications,
etc.

To utilize mm-wave communications for train-ground communications,
high-speed data transmission can be realized with the help of mobile relays (MRs) deployed on the rooftop of the train, which
can
significantly improve
the broadband wireless communication service performance of the entire train-ground communication system~\cite{HSRMR}.
Furthermore, the
full-duplex (FD) communication technology allows wireless communication devices to simultaneously transmit and receive signals on the same frequency band, and the capacity of the communication system can be doubled theoretically~\cite{FDintr}.
Therefore, it
would be desirable for
the MRs
to
adopt both
the mm-wave and FD technology, and simultaneously serve HSR passengers and ground users. The track-side BS allocates part of the bandwidth to the MRs,
allowing
some users
to
directly communicate with the track-side BS in a traditional manner, and
other
users to be
associated
with the MRs to obtain wireless access.
In this system model, we aim to propose an effective algorithm that can fully coordinate the ground BS and MRs to obtain maximum spectrum utilization. Due to the plentiful spectrum resources of mm-wave frequency band, and the HSR scenario has inherent high-speed time-varying characteristics, different bandwidth resource allocation methods may bring great performance differences. Therefore, determining the bandwidth allocation mechanism in the investigated train-ground communication system is a key challenge, and proposing an effective bandwidth resource allocation algorithm to maximize the network capacity is the focus of this paper.

Therefore, it is
timely and relevant
to carry out research on
resource allocation
in
the train-ground communication system based on
mm-wave communications
and FD
transmissions
in HSR scenarios.
Such reseach will
provide theoretical and technical support for the application of mm-wave communication and FD
to a new generation of
HSR
wireless communication systems
and to improve the
quality
of international rail transit communication services.
The current
research efforts on
mm-wave communication is concentrated on 28GHz, 38GHz, 60GHz frequency
bands
and the
E-band (i.e.,
71-76GHz and 81-86GHz). At the same time, the rapid development of Complementary Metal Oxide Semiconductor (CMOS) in radio frequency integrated circuits has paved the way for the production of mm-wave electronic
devices~\cite{Des60GHz}.
Several
international
standardization efforts,
such as ECMA-387~\cite{ECMC}, IEEE 802.15.3c~\cite{80215}, and IEEE 802.11ad~\cite{80211},
focusing on indoor Wireless Personal Area Networks (WPAN) or Wireless Local Area Networks (WLAN),
have made great progress.

In
order to compensate for the
severe
link attenuation of mm-wave communications,
directional antennas
are
used to
utilize the
beamforming technology
for high
antenna gain.
Due to the short wavelength,
many tiny antenna elements can be integrated in a small area.
Now a variety of beam training algorithms have been proposed to
help reduce the time resource required for beam training at the transmitter and receiver~\cite{beamintr}.

Due to
directional transmissions,
the mutual
interference between
mm-wave
links
can be
significantly reduced. Mudumbai \emph{et al.}~\cite{pseudowired}
modeled the
highly directional wireless links in outdoor mesh networks in the 60GHz
band as ``pseudo wired,"
i.e., the details of antenna pattern and interference between the non-adjacent links can be ignored in the MAC protocol design of the mm-wave mesh network. Son \emph{et al.}~\cite{CorTra,CorTra2}
showed
that when the inter-links interference is reduced, multiple data streams can be transmitted in the same time slot to achieve maximizing the space division multiplexing gain.
On the other hand, the directional
transmissions also
make
it
hard
to implement the carrier sense method used by
broadcast channels
to avoid collision, which is called the ``deafness" problem~\cite{CorTra}. Therefore, the design of the MAC protocol in
mm-wave communication scenarios needs to consider the coordination mechanism
of mm-wave links
and make full use of concurrent transmissions
to increase the network capacity.

On the other hand, as a technology with the
great
potential to
increase
the wireless
capacity,
the FD communication is
regarded
as one of the key physical layer technologies of
the
5th generation mobile networks (5G), which has
triggered
great attention
in
academia and industry in recent years~\cite{FD}. However, when the FD technology is adopted, the signal at the transmitting end leaks to the corresponding receiving end to cause
strong self-interference (SI).
At present, the biggest challenge in applying FD technology in different scenarios is to eliminate such
SI.
In order to improve the spectrum efficiency of the FD communication, a variety of physical layer techniques such as antenna interference cancellation, radio frequency interference cancellation and digital interference cancellation have been investigated. Cui \emph{et al.} \cite{FDRelay1} proposed an optimal MR selection scheme in the FD multi-relay communication scenarios to maximize the signal-to-interference plus noise ratio. In order to cope with the SI in the FD Multiple-Input Multiple-Output (MIMO) relay network, Rahman \emph{et al.} \cite{FDRelay2} proposed an efficient SI elimination algorithm based on the Space Projection Algorithm (SPA).
With the latest techniques, the SI level
can be
reduced to -110dB, ensuring that the FD technology can be
applied in practical scenarios.


However, there has been very limited prior work on applying these two advanced technologies, i.e., mm-wave communications and FD, to train-ground communication system in HSR scenarios.
In this paper, we study the bandwidth resource allocation mechanism
for the
mm-wave train-ground communication system
where the MRs adopt FD communications.
We focus on
the coordination of bandwidth resource allocation between track-side BS and FD MRs in
the HSR
scenario,
aiming at
network capacity optimization
for
the train-ground communication system. It is foreseeable that if mm-wave communication and FD
can be adopted
in train-ground communication scenarios, and the bandwidth allocation mechanism developed in this paper
will be useful for
achieving high
global network data transmission rates. The main contributions
made in this paper
are
summarized
as follows.

\begin{itemize}
\item We introduce FD technique to MRs deployed on rooftop of train cabin in mmWave train-ground communication systems. Background noise and residual SI (RSI) are simultaneously taken into account so that the advantages of the FD and mmWave can be fully utilized.

\item By taking the bandwidth allocation ratio of track-side BS and FD MRs as the variable, we formulate the bandwidth allocation problem in the established mmWave train-ground communication systems as a nonlinear programming problem. Then, we take maximizing network capacity as the goal, and propose a fast-converging bandwidth allocation algorithm which can ensure high network capacity and a certain anti-SI ability.

\item We evaluate the proposed algorithm in the 60 GHz mmWave train-ground communication systems with limited spectrum resource. The extensive simulations demonstrate that compared with other traditional priority algorithms and optimization algorithms, the proposed SQP algorithm based on Lagrangian function can significantly increase the network capacity and make the mmWave train-ground communication systems have a certain anti-SI ability. Furthermore, we also give insights into the reasons for the formation of some system performance.
\end{itemize}

The rest of the paper is organized as follows. In Section~\ref{S2}, we
provide
an detailed overview of the related work. In Section~\ref{S3}, we introduce the
train-ground communication system model, and formulate the problem of bandwidth allocation between half-duplex (HD) track-side BS and FD MRs.
The proposed SQP algorithm is presented
in Section~\ref{S4} and evaluated in Section~\ref{S5}.
Finally, Section~\ref{S6} concludes this paper.

\section{Related Work}\label{S2}

We
can classify
the related works into three categories: (a) mm-wave communications in wireless networks; (b) FD communications in wireless networks; and (c) application of optimization theory into resource allocation. We examine these related work in this section.

\paragraph{Mm-wave Communication}
There has been considerable research on mm-wave based wireless communications and networks.
Mm-wave has the advantage of large bandwidth, and
the
short coverage range also makes it a good application in heterogeneous networks. Niu \emph{et al.}~\cite{niuconvey} summarized the characteristics of mm-wave communications,
the research status
in the
physical layer and MAC layer,
and
the
existing problems and challenges,
and discussed the future research directions. Elkashlan \emph{et al.}~\cite{mmwave5G}
showed
that the mm-wave frequency band can provide users with high-capacity wireless broadband access,
e.g.,
a multi-gigabit transmission rate for broadband multimedia services. Niu \emph{et al.}~\cite{niu2017energy} developed an energy-efficient scheme to jointly optimize power control and concurrent transmission scheduling of mm-wave backhauling in small cells
and
densely deployed
heterogeneous cellular networks (HCNs).
A blind neighbor discovery algorithm with bounded discovery time and guaranteed discovery performance was proposed in~\cite{WangICDCS2017}, while a link scheduling algorithm was developed in~\cite{HeAccess2015} for mm-wave ad hoc networks under blockage and interference conditions. However, compared to
sub-6GHz communication systems, mm-wave communications
suffer considerably
greater propagation loss. In practical applications, directional antennas are often used to achieve beamforming to increase antenna gain. In the established mm-wave train-ground communication system model, many directional antennas are
employed.

At present, the research of mm-wave in HSR scenarios has also made progress.
Yang \emph{et al.}~\cite{HSRcha1} and He \emph{et al.}~\cite{HSRcha2} measured the mm-wave propagation in HSR scenarios, and accordingly studied the path loss characteristics.
Based on the path loss measurement at 90 GHz in a railway viaduct scenario, similar calibration works are also performed in~\cite{90GHSRPS}. Gao \emph{et al.}~\cite{HSTdelay} proposed a strategy that can solve the edge caching and content delivery problem for both HSR passengers and low-mobility cellular users.
Wang \emph{et al.}~\cite{energy2} proposed an energy-efficient power-control scheme
for
train-ground mm-wave communications
in HSR scenarios.
However, due to the difficult propagation characteristics of
mm-wave and the particularity of HSR scenarios,
there are still many challenges ahead.

\paragraph{FD Communications}
FD wireless communications have also attracted much efforts in the research community in the past decade.
For example,
Duarte \emph{et al.}~\cite{Duarte} designed a multi-antenna FD physical layer structure and
an
FD MAC layer structure compatible with the existing 802.11
standard.
It turns out that the FD communication mode has
the potential to
nearly doubled the system throughput, and the applications of the FD mode
in
future WiFi standards will have potentially
huge benefits.
The energy-delay trade-off in a multi-channel FD wireless LAN was derived in~\cite{Jiang17} with an application of Lyapunov optimization, while a distributed power control algorithm was developed in~\cite{Wang17} for FD wireless networks by applying the dual-decomposition approach.
Feng \emph{et al.}~\cite{Feng15} showed that the gain of FD depends on the network topology and other settings; its not always the case that FD is better than HD. A joint duplex mode selection, channel allocation, and power control scheme was developed for FD HCNs.
Wen \emph{et al.}~\cite{DingzhuWen} proposed a resource allocation algorithm in a time-division multiplexed cellular network to maximize the throughput of the system while meeting the quality of service requirements of each user as much as possible. In this system, the users' equipments use the traditional HD communication mode, while the BS uses the FD communication mode.

Xiao \emph{et al.}~\cite{Zhenyu} explored the proper antenna configuration for FD mm-wave communications,
and established a related model of the mm-wave SI channel. Ding \emph{et al.}~\cite{dingweiguang} proposed a QoS-aware FD  concurrent transmission algorithm to maximize the number of data streams that meet user
QoS requirements.
Skouroumounis \emph{et al.}~\cite{Skouroumounis} evaluated the impact of the FD
mode in mm-wave communications,
and proposed
an
analytical
framework based on
stochastic
geometry to evaluate heterogeneous FD mm-wave cellular network performance from both cell center users and cell edge users. Anokye \emph{et al.}~\cite{PRINCEANOKYE} established a kind of cellular network topology,
where
the large-scale multiple-input multiple-output (MIMO) system
is applied in the backhaul
to enable FD communications. At the same time, according to the actual number of antennas required and the formed SI thereby, a wireless communication system
with both HD and FD communications
was
proposed.
Wang \emph{et al.}~\cite{YibingWang} studied the optimal allocation of sub-channels under dense deployment of micro-cells in mm-wave networks, and proposed a sub-channel allocation method based on alliance game theory, which improved resource utilization and system throughput.

The related
work show that FD communications
have
great research value in cellular networks, WiFi networks, mm-wave networks, and heterogeneous networks.
However, there are still few applications
of FD
in HSR scenarios. In the train-ground communication system model established
in this paper,
we
exploit
the advantages of FD technology and mm-wave communications,
which can
jointly
bring about
huge improvements
on
spectrum efficiency and network capacity.

\paragraph{Optimization Resource Allocation}
Optimization theory has been widely applied in wireless networks to improve the utilization of limited resources, such as spectrum and power.
For instance,
Chen \emph{et al.}~\cite{chen2018allocation} proposed a coalition formation game to optimize resource allocation to maximize the system sum rate of HCNs in
the
statistical average sense. Liu \emph{et al.}~\cite{AntOptimization} studied the sub-channel allocation problem in
two-tier
OFDMA
femtocell networks, and proposed a resource optimization method based on
ant colony optimization
to maximize the rate of multiple femtocells systems,
while considering the
cross-tier
interference between the macrocell and multiple femtocells. Xiao \emph{et al.}~\cite{OptimalXiao} used robust optimization theory to model uncertain interference channels and considered channel estimation errors to maximize the
network
throughput
while avoiding the
severe
inter-tier
interference.
With
the Lagrangian dual method, the original optimization problem
was
decomposed into an original problem and a dual problem
to be solved.

Motivated by these related works, we
explore
optimization theory to model the bandwidth allocation problem of BS and MRs in
the
train-ground communication system. The key
is to introduce a suitable algorithm to maximize the network capacity.
The proposed algorithm is designed
to solve the bandwidth allocation problem. It is also hoped that it has a certain degree of anti-SI ability
for train-ground communication
applications.

\section{System Overview and Problem Formulation}\label{S3}

\subsection{System Model}\label{S3-1}

\begin{figure}[!t]
	\begin{center}
		 \includegraphics*[width=1\columnwidth,height=1.85in]{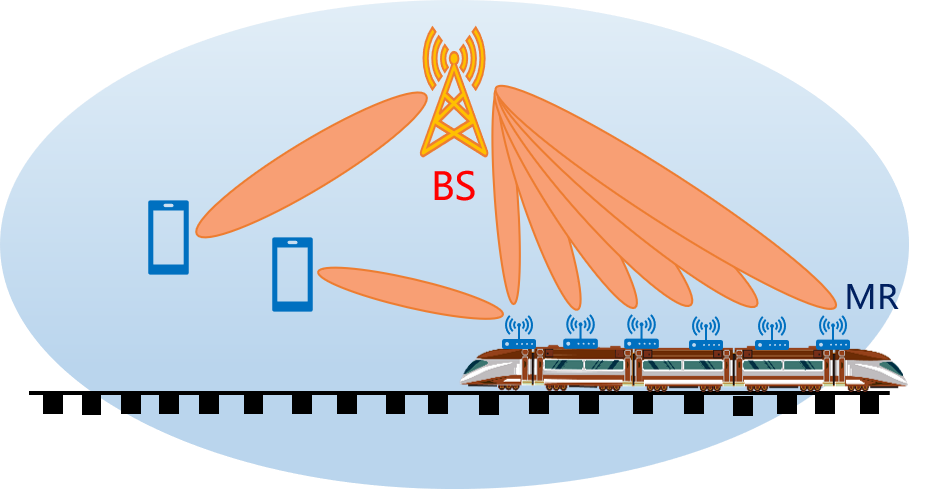}
	\end{center}
	\caption{Illustration of an mm-wave train-ground communication system
	using multiple FD MRs.} \label{fig3-1}
\end{figure}

This paper considers an
mm-wave train-ground communication system using FD MRs,
as
shown in Fig.~\ref{fig3-1}.
In this model,
the track-side BS reserves part of the bandwidth, and the remaining bandwidth is allocated to the MRs for FD communications.
In addition,
there are multiple
MRs
operating in the FD mode
deployed
on the train.
The bandwidth resources allocated to each MR are mutually exclusive.
All the devices in
this train-ground communication system work in the mm-wave frequency band. Both the BS and the MRs are equipped with
steerable
directional antennas,
which
allow the BS
and
the MRs to
aim
at their associated users to
achieve
higher antenna gain.

The track-side BS and each MR are associated with different users, and it is
assumed
that these association relationships have been determined in advance. Assume that multiple served users
access the BS or a specific MR with
Time Division Multiplexing Access (TDMA),
and
the
time slot resources are equally allocated. In the following, for this mm-wave train-ground communication system using
FD MRs, it is considered that the total initial system bandwidth, i.e., the sum of the bandwidth of the BS and the bandwidth allocated to all MRs, is constant.
We
hope to find the best allocation ratio of bandwidth resources to
achieve high
network capacity. When the train is stationary, the global network capacity
of
this system is defined as the sum of the average user rates
at the BS and
that at all the
MRs.

Since the
train-ground communication network works in the mm-wave frequency band, non-line-of-sight transmissions
will cause greater attenuation. Considering that both the BS and the MRs are equipped with
steerable
directional antennas, this paper assumes that all
the nodes in
the system use
the line-of-sight
links
for transmission.
According to
the path loss model in~\cite{PathLossModel}, the received power
at
receiver ${r_k}$ from its associated server transmitter ${t_s}$ can be expressed as
\begin{align}
{P_r}(s,k) = {k_0} \cdot {G_t}(s,k)  \cdot {G_r}(s,k) \cdot l_{s \to k}^{ - n} \cdot {P_t},
\label{eq3-2}
\end{align}
where
${k_0}$ is a constant coefficient, which is proportional to ${(\frac{\lambda}{{4\pi}})^2}$,
$\lambda$ is the wavelength of the carrier frequency,
$n$ is the path loss index, ${P_t}$ represents the transmit power of the transmitter ${t_s}$, ${G_t}(s,k)$ is the transmit antenna gain of
transmitter ${t_s}$ in the direction of
user $k$'s
receiver ${r_k}$,
${G_r}(s,k)$ represents the
receive
antenna gain from the server transmitter ${t_s}$ to the user $k$'s
receiver ${r_k}$, ${l_{s \to k}}$ indicates the distance between 
user $k$'s
receiver ${r_k}$ and the server transmitter ${t_s}$.

Due to the inherent multipath effect of
wireless channels, the mm-wave channel is modeled as a Gaussian channel. According to the Shannon channel capacity formula, The data transmission rate ${R_s}(k)$ obtained by
user $k$, who accesses the network through
service device $s$, can be expressed as
\begin{align}
{R_s}(k) = \eta \cdot {W_s} \cdot {\log_2} \left( 1 + \frac{{{P_r}(s,k)}}{{{N_0}{W_s}}} \right),
\label{eq3-3}
\end{align}
where
$\eta \in (0,1)$ characterizes the
efficiency of the signal transceiver, ${W_s}$ represents the bandwidth resource reserved
for
the service device $s$
according to
the bandwidth allocation decision, and ${N_0}$ is the unilateral noise power spectral density
of
the Gaussian channel.

\subsection{Problem Formulation}\label{S3-2}

Let the total system bandwidth be $W$
and the bandwidth reserved by the track-side BS
be
denoted
by
${W_{BS}}$. We introduce a bandwidth resource allocation factor vector
$\boldsymbol{\alpha}$, the
length of which
is the sum of the numbers of MRs and BS in the system. The
following formulation
is based on the scenario of a
single
BS and multiple MRs,
which, however, can be easily extended to the more general cases.
The first element ${\alpha_0}$ of ${\boldsymbol{\alpha}}$ is the ratio of the bandwidth resources allocated to the BS,
i.e., ${W_{BS}}$,
over the total bandwidth $W$.
The remaining
elements ${\alpha_i}$ represents the ratio of the bandwidth allocated to
MR $i$, i.e., ${W_i}$,
over $W$. The related mathematical relations of bandwidth resources in the system can be expressed
as in~(\ref{eq3-567}).
\begin{align}
\left\{ \begin{array}{l}
W = {W_{BS}} + \sum\limits_{i = 1}^{Rnum} {{W_i}} \\
{{W_{BS}} = W \cdot {\alpha _0}} \\
{{W_i} = W \cdot {\alpha _i}}, \;\; i=1, 2, ..., R_{num},
        \end{array} \right. \label{eq3-567}
\end{align}
where $Rnum$ represents the total number of MRs in the HSR scenario.

The total number of users in the train-ground communication system is denoted as $M$, the number of users associated with the BS is denoted as $M_{BS}$, and the number of users associated with
MR $i$
is denoted as $M_i$. Using the received power model~(\ref{eq3-2}),
the received power by user $k$ associated with the BS (with transmit power ${P_t}(BS)$) can be expressed as
\begin{align}
{P_r}(BS,k) = {k_0} \cdot {G_t}(BS,k) \cdot {G_r}(BS,k) \cdot l_{BS \to k}^{-n} \cdot {P_t}(BS).
\label{eq3-8}
\end{align}
Similarly,
for the $M_i$ users associated with
MR $i$,
the received power
by user $j$ from MR $i$ (with transmit power ${P_t}(i)$) can be expressed as
\begin{align}
{P_r}(i,j) = {k_0} \cdot {G_t}(i,j) \cdot {G_r}(i,j) \cdot l_{i \to j}^{-n} \cdot {P_t}(i).
\label{eq3-9}
\end{align}

The data transmission rate of
user $k$
associated with the BS is denoted as ${R_{BS}}(k)$, which can be expressed as
\begin{align}
{R_{BS}}(k) = \eta \cdot {W_{BS}} \cdot {\log_2} \left(1 + \frac{{{P_r}(BS,k)}}{{{N_0}{W_{BS}}}} \right).
\label{eq3-10}
\end{align}
%
Since the BS uses TDMA for
associated users, the average transmission rate of the users associated with the BS can be expressed as
\begin{align}
 \bar{R}_{BS} = \frac{1}{M_{BS}} \sum_{k = 1}^{M_{BS}} R_{BS}(k).
\label{eq3-11}
\end{align}

\begin{figure}[!t]
	\begin{center}
		 \includegraphics*[width=0.6\columnwidth,height=1.33in]{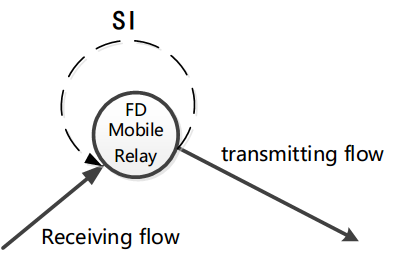}
	\end{center}
	\caption{SI at an FD MR.} \label{fig3-2}
\end{figure}

Similarly, the rate of
user
$j$
associated with
MR
$i$
is denoted as ${R_i}(j)$, which can be expressed as
\begin{align}
{R_i}(j) = \eta \cdot {W_i} \cdot {\log_2} \left( 1 + \frac{{{P_r}(i,j)}}{{{N_0}{W_i} + {I_s}(i)}} \right).
\label{eq3-12}
\end{align}
It should be noted that, because the MRs
operates in the FD mode,
the signal from a certain MR will suffer from
the background noise
as well as a certain amount of SI (i.e., due to imperfect SI cancellation).
As shown in Fig.~\ref{fig3-2}, the so-called SI refers to the interference
at the MR receiver caused by its own transmitter.
We introduce
\begin{align}
	{I_s}(i) = \beta \cdot {P_t}(i),
\end{align}
to
model
the residual SI level
at MR $i$,
where $\beta$ is the SI cancellation level.
The smaller the $\beta$, the
more effective the SI cancellation at the MR.
Since the MRs also use TDMA for the data
transmissions of
their
associated users, the average transmission rate of the users associated with
MR
$i$
can be expressed as
\begin{align}
\bar{R}_i = \frac{1}{M_i} {\sum_{j = 1}^{{M_i}} {{R_i}(j)} }.
\label{eq3-13}
\end{align}

According to the
train-ground communication system model,
we next formulate the mathematical model for
the goal of maximizing network capacity.
With the models given in Section~\ref{S3-1} and this section, the objective function of the
bandwidth allocation problem can be expressed as
\begin{align}
\max \;\; \bar{R}_{BS}  + \sum_{i = 1}^{Rnum} {\bar{R}_i},
\label{eq3-1}
\end{align}
where $\bar{R}_{BS}$ is the average
data transmission rate
of the users
associated with the BS, and $\bar{R}_i$ represents the average
data transmission rate
of the users
associated with MR $i$.

Aiming at maximizing the global network capacity of the train-ground communication system, we have established a mathematical model that takes multi-dimensional bandwidth allocation factor ${\boldsymbol{\alpha}}$ as a variable. The objective function is a multivariate nonlinear
function.
%
The
constraints
on
the bandwidth allocation factor ${\boldsymbol{\alpha}}$
in practical
HSR scenarios
can be expressed as
\begin{align}
& \alpha_0 + \sum_{i=1}^{Rnum} \alpha_i = 1
\label{eq3-14} \\
& 0 \le {\alpha_i} \le 1, \ \ i = 0, 1, ..., Rnum.
\label{eq3-15}
\end{align}

Due to the short wavelength, mm-wave is vulnerable to various blockages, such as foliage, buildings, and viaducts in railway environments. The link blockage occurs when obstacles appear in the radio links between the transceivers, which results in received signal strength degradation caused by severe attenuation~\cite{mmwaveblockage}. When the achieved signal to interference plus noise ratio (SINR) at the receiver (RX) side is lower than the required threshold ${\gamma _{th}}$, the system is unable to guarantee the required bit error rate and the link is considered to be ¡°turned off¡±~\cite{turnoff}. The link blockage depends on multiple factors, including the surrounding environment, obstacle density, beamwidth, and transmission distance.

Considering the height of the MRs, the link between the MRs and the ground BS can be regarded as line-of-sight (LOS) transmission, and the railway is generally located in spacious scenarios, so it is considered that the blockages are mainly caused by human body. We consider that in the investigated mm-wave train-ground communication system, the blockage problem caused by human body can be solved in three steps. In the first step, we connect the users to the nearest BS or MR, and propose an average outage probability ${P_b}$ for all links, which note that each link has a random probability of experiencing blockage. In the second step, if the communication link between the user and the associated equipment experiences blockage, this user will re-establish an association relationship with the second closest equipment to actively avoid blockage. In the third step, we study the scenario of non-LOS (NLOS) transmission, which can be enhanced with the assist of intelligent reflecting surface (IRS) and Device-to-Device (D2D) communications. In this paper, for analytical simplicity, it is assumed that the link blockage probability remains stable in a road segment. The link blockage probability ${P_b}$ on average is deemed as a constant in a certain section along the rail track~\cite{blockconstant}.

\begin{figure}[!t]
	\begin{center}
		 \includegraphics*[width=1\columnwidth,height=1.4in]{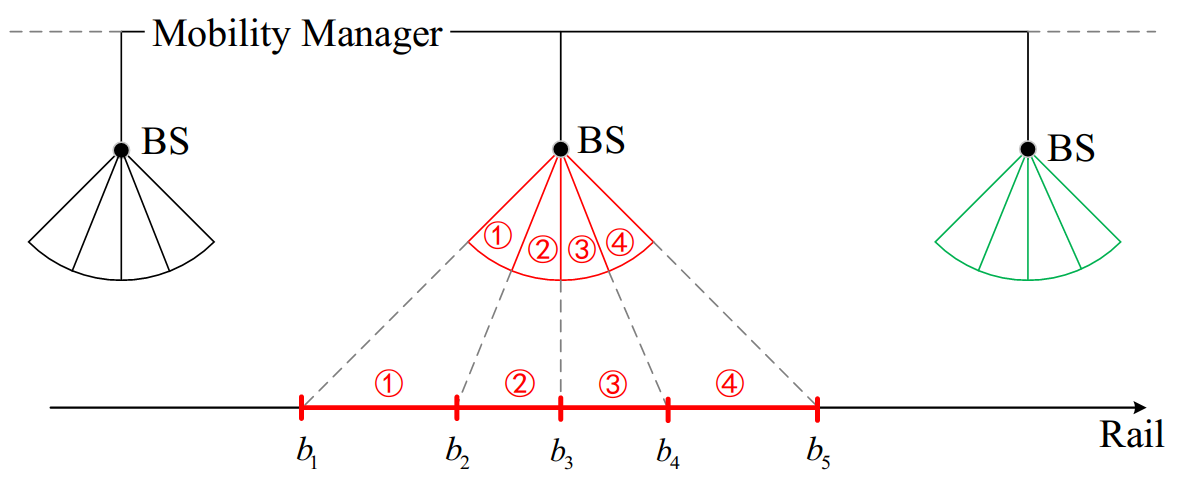}
	\end{center}
	\caption{Beam switching to get a LOS link.} \label{figbeam}
\end{figure}

In order to construct a LOS link in the investigated mm-wave train-ground wireless communications, we consider adopting the beam switching for beam alignment. The mm-wave BS communicates with the MRs using directional beamforming at the TX and RX. Train control systems (TCSs) are used in modern railway systems, which can help us obtain the train position and speed information in time and achieve efficient beam alignment. Each beam can be projected on to the rail as shown in Fig.~\ref{figbeam}, which defines its coverage~\cite{beamalign}. If it is known that MR have entered the coverage of a certain beam, then this beam is the best choice of the MR for mm-wave communications with BS. Since the speed of the high-speed train is relatively stable during the driving process, this beam alignment approach using train position information avoids the cumbersome traditional beam training process~\cite{Liyan}.

\begin{figure}[t]
\begin{minipage}[t]{0.5\linewidth}
\centering
\includegraphics[width=1\columnwidth,height=1.3in]{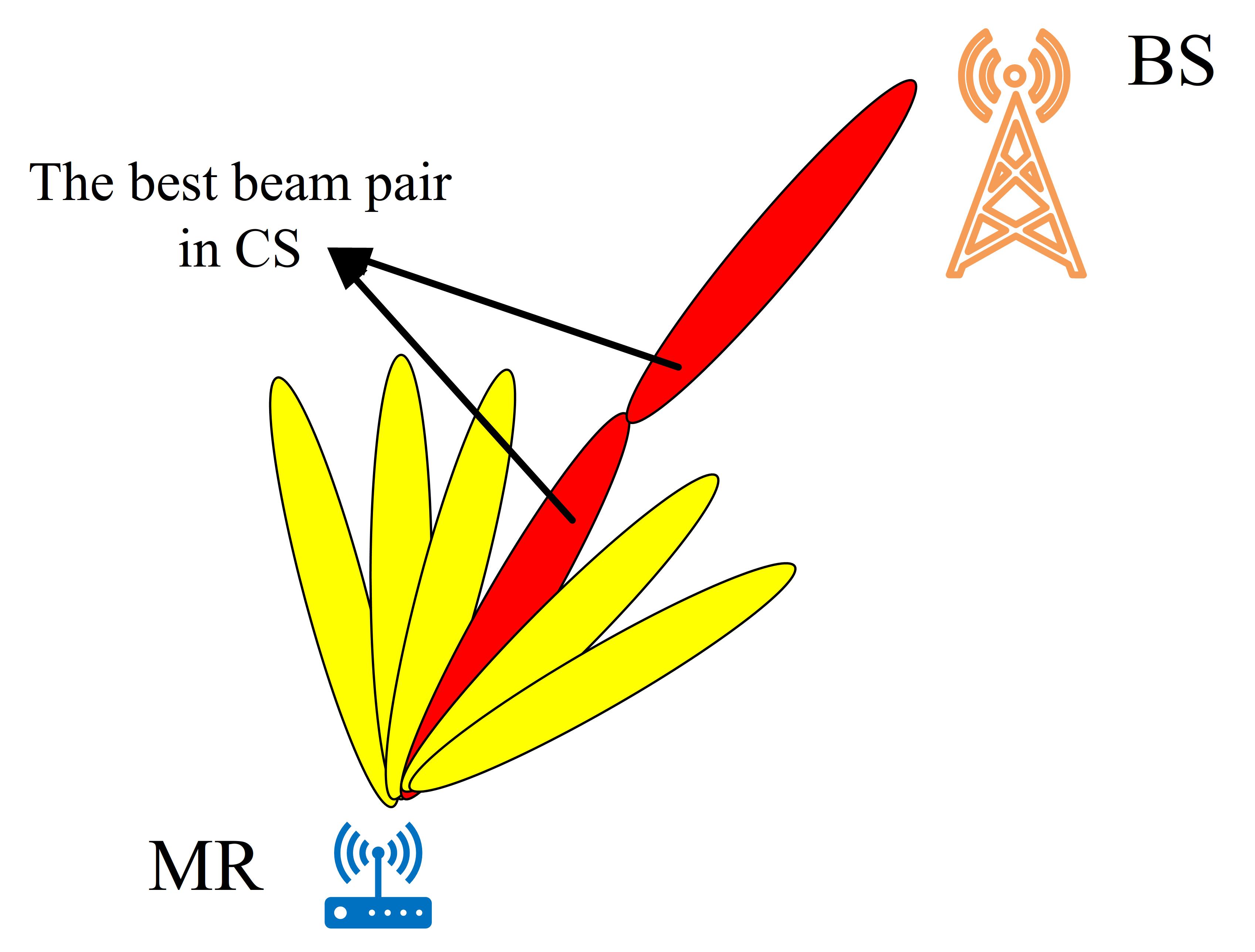}
\centerline{\small (a)}
\end{minipage}%
\begin{minipage}[t]{0.5\linewidth}
\centering
\includegraphics[width=1\columnwidth,height=1.3in]{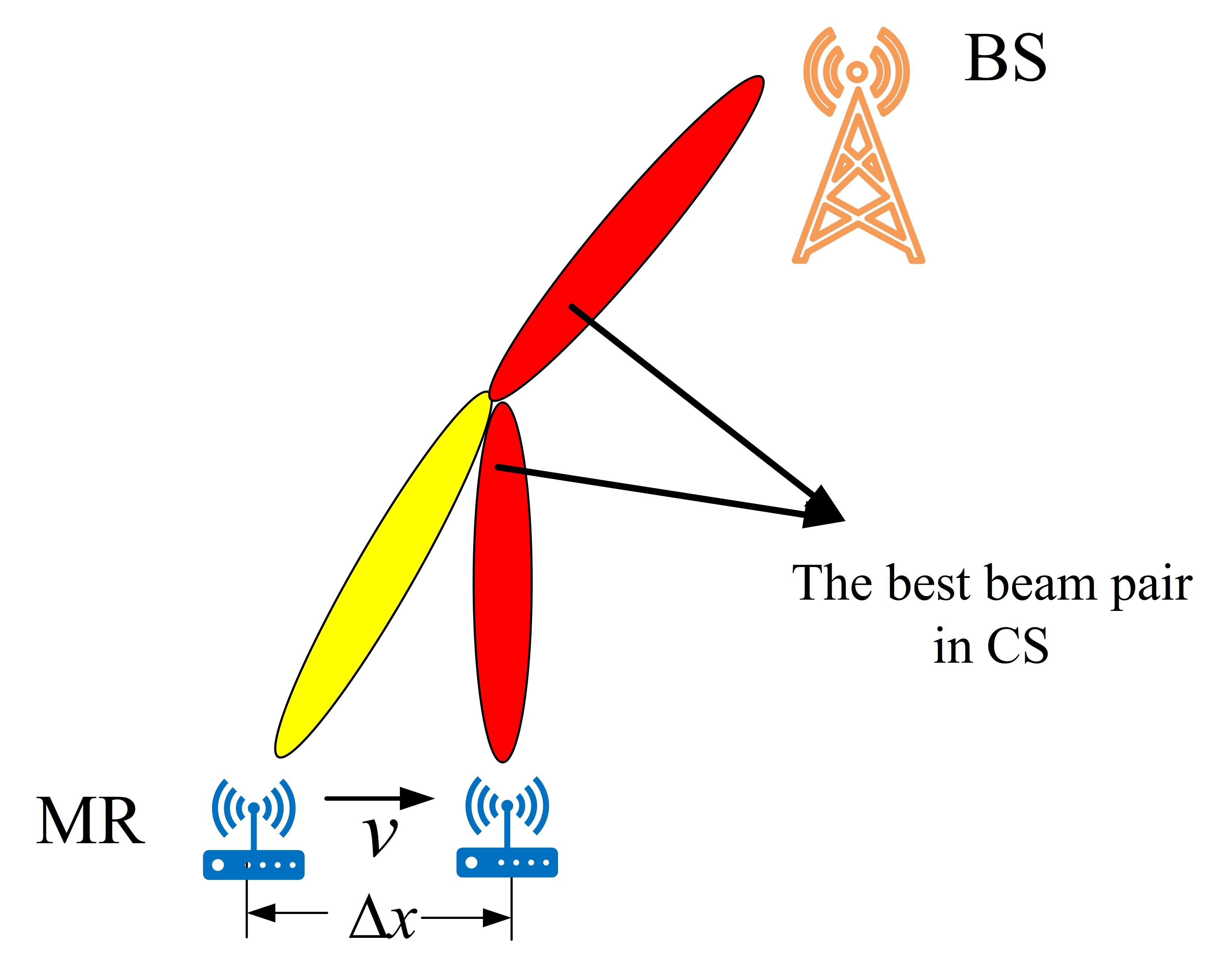}
\centerline{\small (b)}
\end{minipage}
\caption{The IA process: (a) CS phase, (b) RA phase.}
\vspace*{-3mm}
\label{figCSRA}
\end{figure}

In current mm-wave systems, the entire initial access (IA) process consists of two parts, the cell search (CS) and random access (RA) phases~\cite{CS}\cite{RA}, and beam switching avoids waste caused by beam scanning. As shown in Fig.~\ref{figCSRA}(a), in the CS phase, MRs measure the signal quality of each beam pair according to all Feasible direction in the beam coverage area, and record the index of the best transmitting beam. In the following RA phase as shown in Fig.~\ref{figCSRA}(b), at first users randomly select a preamble from a predefined preamble set in which the preambles are mutually orthogonal~\cite{preamble}. Usually, the number of orthogonal preambles determines the maximum number of MRs to be allowed to simultaneously access this network. By coding the recorded index of the best transmitting beam with the selected preamble, the user then sends this information through the exact receiving direction from which the best transmitting beam is detected during the previous CS phase. Meanwhile, the BS can capture this feedback~\cite{Liyan}.

Besides, in mm-wave systems, the preamble-coded information of the RA phase is carried by narrow beams and transmitted only to the direction where the best transmitting beam is detected in the CS phase, thereby highly decreasing the access collision probability compared with omni-directional wireless systems~\cite{CS}.

In summary,
the problem of optimal bandwidth allocation, denoted by P1, can be formulated as follows.
\begin{align}
\mbox{(P1)} \;\;
\max &~~(1-P_b)(\bar{R}_{BS} + \sum_{i = 1}^{Rnum} {\bar{R}_i}) \label{P1}
\\
\mbox{s.t.} &~~\mbox{Constraints~(\ref{eq3-14}) and~(\ref{eq3-15})}. \nonumber
\end{align}

Because
the data transmission rate of each user includes a logarithmic function,
the objective function is nonlinear.
So problem (P1) is a nonlinear programming problem.
In the next section,
we introduce
a solution
algorithm to
solve problem (P1) with competitive solutions.

\section{Bandwidth Allocation Algorithm}\label{S4}

In this
section,
we will introduce a sequential quadratic programming (SQP) algorithm based on Lagrangian function for solving (P1). With the help of the Lagrangian function, the
proposed
algorithm ensures that the approximated sub-problem of the original problem is always a convex optimization problem, and the SQP algorithm ensures the speed of
convergence to
the optimal solution.
In the following,
we first describe how to construct the Lagrangian function, and then
present
the
SQP algorithm in detail.

\subsection{Lagrange Function}\label{S4-1}

The traditional Lagrangian multiplier method first constructs a Lagrangian function, and then obtains the partial derivative
with respect to
each variable and the Lagrange multipliers.
Letting
the partial derivatives be equal to 0,
finally the extreme values of the problem
can be
obtained.
In problem (P1),
according to the constraint conditions for the bandwidth allocation factor ${\boldsymbol{\alpha}}$
in~(\ref{eq3-14}) and~(\ref{eq3-15}), it can be seen that these contains include an
equality
condition and several inequality conditions. According to~\cite{Lag}, the Lagrangian multiplier method can
effectively
solve the equality constraint problem and the inequality constraint problem.

The Lagrangian function of
problem (P1) is given by
\begin{align}
\mathcal{L}(\boldsymbol{\alpha}, \mu, \boldsymbol{\gamma}) =&\; \bar{R}_{BS} + \sum_{i = 1}^{Rnum} \bar{R}_i + \mu \left( {\alpha_0} + \sum_{i = 1}^{Rnum} \alpha_i - 1 \right) \nonumber \\
&\; + \sum_{j = 0}^{2(Rnum + 1)} \sum_{i = 1}^{Rnum} {{\gamma_j} \cdot {g_j}(\alpha_i)}, \label{eq3-16}
\end{align}%
where
$\mu$ and $\boldsymbol{\gamma}$ are Lagrange multipliers,
the sum of the first two items on the right side is
the objective function of problem (P1), i.e., the network capacity.
The third term takes into account the equality constraint
in~(\ref{eq3-14}). The fourth term takes into account the inequality constraints in~(\ref{eq3-15}), the specific form is to write the constraints ${\alpha_i} \in (0,1)$, $i=0, 1, ..., R_{num}$ in~(\ref{eq3-15}) as two inequalities, which must be in accordance with the form ${g_j}(\alpha_j) \le 0$. Take the bandwidth allocation factor ${\alpha_0}$ of the BS as an example, the two inequality constraints can be expressed as
\begin{align}
\left\{ {\begin{array}{*{20}{l}}
g_0(\alpha_0) = {{\alpha_0} - 1 \le 0}\\
g_1(\alpha_0) = { - {\alpha _0} \le 0}.
\end{array}} \right.
\label{eq3-17}
\end{align}
Now, the actual conditions $0 \le {\alpha _i} \le 1$ that each bandwidth allocation factor needs to meet can be obtained as in~\eqref{eq3-17}.

In addition, the feasible solution obtained by using
the
Lagrangian function
involving
both
equality and inequality constraints needs to meet the Karush-Kuhn-Tucker (KKT) condition~\cite{Lag}.
For problem (P1),
the KKT condition of the corresponding Lagrangian function can be expressed as
\begin{align}
\left\{ \begin{array}{*{20}{l}}
\nabla_{\boldsymbol \alpha} \mathcal{L} (\boldsymbol{\alpha}, \mu, \boldsymbol{\gamma}) = 0 \\
\gamma_j \cdot {g_j}(\alpha_j) = 0, \ \  j = 0, 1, ..., 2(Rnum + 1) \\
g_j (\alpha_j) \le 0, \ \ j = 0, 1, ..., 2(Rnum + 1) \\
\alpha_0 + \sum_{i=1}^{Rnum} \alpha_i - 1 = 0 \\
\gamma_j \ge 0, \ \ j = 0, 1, ..., 2(Rnum + 1).
\end{array} \right.
\label{eq3-18}
\end{align}
In~\eqref{eq3-18},
the first
equation
represents the use of the constructed Lagrangian function to obtain the partial derivative
with respect to
the bandwidth allocation variable, which is also a necessary condition for the Lagrangian multiplier method to obtain a feasible solution.
The second
equation
is
the
Lagrangian slack complementary condition,
while
the third and
fourth
lines are the initial constraints that need to be met in
problem (P1).
And the
fifth line is
the extension under the existence of inequality constraints. These are the conditions that need to be met
by the feasible solution obtained by the Lagrangian multiplier method.
Each solution obtained by the Lagrange multiplier method will be
judged, and the solution satisfying the KKT conditions is the optimal solution
to
the original optimization problem.

It can be
shown
that
the
solution that satisfies the KKT conditions,
obtained
by constructing a Lagrangian function,
is also a feasible solution
to problem (P1).
Before verification,
recall the
mm-wave train-ground communication system model using the FD MRs. Assuming that there are one track-side BS and
nine
MRs in the system model, then the dimension of the bandwidth allocation vector ${\boldsymbol{\alpha}}$ is 10, and the objective function is a non-linear function of 10 variables.
The overall constraints include one equation and 20 inequalities,
and
then the partial derivatives of the Lagrangian function contain 31 variables.
Solving the system of nonlinear equations will be highly complicated when the system is large.
Therefore, we consider using the optimization algorithm to perform multiple iterations to quickly approximate the optimal solution.
Dynamic adjustment of the convergence accuracy can also increase the calculation speed or accuracy, and it provides greater flexibility for
the real
communication scenarios to deal with real-time service
requirements.

\subsection{SQP Algorithm}\label{S4-2}

The SQP
method
transforms the originally complicated nonlinear programming problem into a series of quadratic programming sub-problems.
In
the beginning,
an
approximation
solution
is used to approximate the objective function of the original
problem,
which
is simplified into a quadratic programming sub-problem.
And
the approximation
solution of the original
problem is obtained
with
this quadratic programming sub-problem. If the convergence accuracy is met, the obtained
approximation solution
is considered to be the optimal solution
to
the original problem.
Otherwise,
the starting point is selected again to repeat the above
procedure.
This is
an
iterative calculation process
to solve the complicated nonlinear programming problem.

Compared with other optimization algorithms, SQP
has the outstanding advantages of fast convergence, high computational efficiency,
and strong boundary search ability, and has been widely used.
On the other hand, in order to
achieve
global convergence, it is usually required that the second derivative matrix $H$ of the objective function of the quadratic programming sub-problem is symmetric and positive definite. At this point, this sub-problem can be further characterized as a strict convex quadratic programming problem.
This type of problems
have
a unique solution, i.e., the local optimal solution is
also
the global optimal solution, and it is relatively easy to obtain this unique solution.
Moreover,
the positive definiteness of
matrix $H$ ensures that the search direction obtained by solving the quadratic programming sub-problem is indeed the decreasing direction of the objective function value of the original problem. The specific implementation principle of the SQP algorithm
is given
below.

Consider
a
nonlinear programming problem that includes
both
equality and inequality constraints
given by~(\ref{eq3-192021}).
\begin{align}
\min &\;\; f(\mathbf X) \label{eq3-192021} \\
\mbox{s.t.} &\;\; 
{{g_u}(\mathbf X) \le 0, \ \ u = 1, 2, ..., p} \nonumber \\
&\;\; {{h_v}(\mathbf X) = 0, \ \ v = 1, 2, ..., m}, \nonumber
\end{align}
Obviously,
where $\mathbf X$ represents multi-dimensional variables. 
Then,
applying
Taylor
series
expansion,
the objective function of the original nonlinear programming problem
can be
approximated by the first- and second-order terms (i.e., by discarding higher-order terms) at the iteration point ${\mathbf{X}_k}$,
while the constraint functions
are approximated by their first-order terms.
This way, the original
problem is simplified into a quadratic programming sub-problem
as given in~(\ref{eq3-222324}).
\begin{align}
\min &\;\; f(\mathbf X) = \frac{1}{2}{[\mathbf{X} - {\mathbf{X}_k}]^{\rm T}}{\nabla ^2}f({{\mathbf X}_k})[\mathbf{X} - {{\mathbf X}_k}] \label{eq3-222324} \\
  & \quad\quad\quad\quad + \nabla f{({{\mathbf X}_k})^{\rm T}}[{\mathbf X} - {{\mathbf X}_k}] \nonumber \\
\mbox{s.t.} &\;\; {\nabla {g_u}{{({{\mathbf X}_k})}^{\rm T}}[{\mathbf X} - {{\mathbf X}_k}] + {g_u}({{\mathbf X}_k}) \le 0, \ \ u = 1, 2, ... p} \nonumber \\
&\;\; {\nabla {h_v}{{({{\mathbf X}_k})}^{\rm T}}[{\mathbf X} - {{\mathbf X}_k}] + {h_v}({{\mathbf X}_k}) = 0, \ \ v = 1, 2, ..., m}. \nonumber
\end{align}

It is worth noting that the constrained optimization problem~(\ref{eq3-222324})
is an
approximation
of the original
problem~\eqref{eq3-192021}, but its solution is not necessarily a feasible solution
to
the original
problem. Therefore, we define a new problem variable $\mathbf S$, and its mathematical relationship with the original variable $\mathbf X$ can be expressed as
\begin{align}
\mathbf{S} = \mathbf{X} - \mathbf{X}_k.
\label{eq3-25}
\end{align}
With~(\ref{eq3-25}), we can
rewrite
the simplified quadratic programming sub-problem as a nonlinear programming problem with variable $\mathbf S$,
given by
\begin{align}
\min &\;\; f(\mathbf{S}) = \frac{1}{2}{\mathbf{S}^{\rm T}}{\nabla ^2}f({{\mathbf X}_k})\mathbf{S} + \nabla f{({{\mathbf X}_k})^{\rm T}}\mathbf{S} \label{eq3-262728} \\
\mbox{s.t.} &\;\; {\nabla {g_u}{{({{\mathbf X}_k})}^{\rm T}}\mathbf{S} + {g_u}({{\mathbf X}_k}) \le 0, \; u = 1, 2, ..., p} \nonumber \\
&\;\; {\nabla {h_v}{{({{\mathbf X}_k})}^{\rm T}}\mathbf{S} + {h_v}({\mathbf{X}_k}) = 0, v = 1, 2, ..., m}. \nonumber
\end{align}
We also define the following for~(\ref{eq3-262728}).
\begin{align}
\left\{ \begin{array}{l}
\mathbf{H} = {\nabla ^2}f({\mathbf{X}_k})\\
\mathbf{C} = \nabla f({\mathbf{X}_k})\\
\mathbf{A} = {[\nabla {g_1}({{\mathbf X}_k}), \nabla {g_2}({{\mathbf X}_k}), ..., \nabla {g_p}({{\mathbf X}_k})]^{\rm T}} \\
{{\mathbf A}_{eq}} = {[\nabla {h_1}({{\mathbf X}_k}), \nabla {h_2}({{\mathbf X}_k}), ..., \nabla {h_m}({{\mathbf X}_k})]^{\rm T}}\\
{\mathbf B} = {[{g_1}({{\mathbf X}_k}), {g_2}({{\mathbf X}_k}), ..., {g_p}({{\mathbf X}_k})]^{\rm T}} \\
{{\mathbf B}_{eq}} = {[{h_1}({{\mathbf X}_k}), {h_2}({{\mathbf X}_k}), ..., {h_m}({{\mathbf X}_k})]^{\rm T}}.
\end{array} \right.
\label{eq3-29-34}
\end{align}
This way, the general form of the quadratic programming problem is formed, which is given by
\begin{align}
\min &\;\; \frac{1}{2}{{\mathbf S}^{\rm T}} \mathbf{HS} + {{\mathbf C}^{\rm T}}{\mathbf S} \label{eq3-353637} \\
\mbox{s.t.} &\;\; {\mathbf{AS} \le {\mathbf B}} \nonumber \\
&\;\; {{{\mathbf A}_{eq}} {\mathbf S} = {{\mathbf B}_{eq}}}. \nonumber
\end{align}

To solve the
quadratic programming sub-problem~\eqref{eq3-353637}, the optimal solution ${\mathbf S}^*$ is taken as the next search direction of the original optimization problem, and a constrained one-dimensional search will be performed on the original optimization problem in this search direction. The one-dimensional search mentioned here is also called linear search, which refers to the optimization problems of single-variable function, and it is the basis of multi-variable function optimization problems. The common dichotomy and interpolation are typical one-dimensional search algorithms. The solution obtained based on the constrained one-dimensional search can be regarded as
an
approximation
solution $\mathbf{X}_{k + 1}$ of the original
problem. Iterative calculation of the above process can obtain the optimal solution
to
the original problem within the
prescribed
convergence accuracy. The key
of implementing the SQP
method
based on the above ideas is how to quickly calculate the second derivative matrix $\mathbf H$ of the objective function.

Mathematicians Polomares and Mangasarian proposed a calculation method of $\mathbf H$ in 1976, using the Hessian matrix of the Lagrangian function constructed based on the original optimization problem,
to achieve continuous correction of the second derivative matrix $\mathbf H$ of the objective function of the original
problem~\cite{Palomares1976}.
The Hessian matrix here refers to a matrix composed of the arrangement of the second-order partial derivatives of a multivariate function. Simply put, each iteration performed by the SQP
will make the matrix $\mathbf H$ of the quadratic programming sub-problem closer to the second-order derivative matrix of the objective function of the original
problem. The specific iterative updates to $\mathbf H$
is given by
\begin{align}
& {\mathbf{H}_{k + 1}} = {\mathbf{H}_k} + \frac{{{\mathbf{q}_k}{\mathbf{q}_k}^{\rm T}}}{{{\mathbf{q}_k}^{\rm T}{\mathbf{S}_k}}} - \frac{{{\mathbf{H}_k}^{\rm T}{\mathbf{H}_k}}}{{{\mathbf{S}_k}^T{\mathbf{H}_k}{\mathbf{S}_k}}}\label{eqBFGS} \\
& {{\mathbf{S}_k} = \mathbf{X} - {\mathbf{X}_k}} \\
& {{\mathbf{q}_k} = \nabla f({\mathbf{X}_{k + 1}}) + \sum\limits_{u = 1}^p {{\gamma _u}\nabla {g_u}({\mathbf{X}_{k + 1}}) + \sum\limits_{v = 1}^m {{\mu _v}} } \nabla {h_v}({\mathbf{X}_{k + 1}})} \nonumber \\
& {\ \ \ \ - [\nabla f({\mathbf{X}_k}) + \sum\limits_{u = 1}^p {{\gamma_u}\nabla {g_u}({\mathbf{X}_k}) + \sum\limits_{v = 1}^m {{\mu _v}} } \nabla {h_v}({\mathbf{X}_k})]}.
\label{eq3-383940}
\end{align}
In the iterative process of the SQP algorithm, as long as it is guaranteed that
$\mathbf{q}_k^{\rm T}{\mathbf{S}_k}$
is positive and $\mathbf{H}$ is initialized to a positive definite matrix, the Hessian matrix will always remain positive definite.~(\ref{eqBFGS}) is called BFGS correction method.
Therefore, the key problem of
applying the
SQP algorithm to solve problem (P1) has been solved.

Sorting out the basic ideas of the SQP algorithm, combined with
problem (P1)
studied in this paper, we
find that the objective function of
problem (P1) is the
function
each time the SQP algorithm performs a secondary approximation, and the so-called ``secondary approximation" is actually a second-order Taylor expansion of the objective function, but each iteration is carried out at a different operating
point.
Section~\ref{S4-1} describes the process of constructing a Lagrangian function
for problem (P1),
and this Lagrangian function is used to modify the positive definite matrix $\mathbf H$, which is needed every time the SQP algorithm simplifies the original complicated nonlinear programming problem.
Quadratic programming is a natural transition from linear programming to nonlinear programming.
Finally, by repeatedly determining the search direction and performing iterative calculations, the optimal solution to
problem (P1) can be obtained
when
the convergence accuracy
condition is satisfied.

The pseudo code
of the SQP algorithm
is
given
in Algorithm~\ref{alg1},
which is executed at the {\em centralized controller} (CC) located
in
the backbone network.
It can be seen from Algorithm~\ref{alg1}
that the Lagrangian function is constructed by sorting out the objective function and
the
related constraints of the original bandwidth allocation problem.
Then
the SQP algorithm is used to simplify the complicated nonlinear programming problem
to a series of quadratic programming sub-problems.
In the
iterative calculation process, the constructed Lagrangian function is used to modify the second derivative matrix $\mathbf H$ of the original objective function. By solving the quadratic programming sub-problems, judging the convergence accuracy, and iterative calculating, the
approximation
solution to the original problem (P1) is obtained.

The computational complexity analysis of the proposed Algorithm 1 is crucial and necessary. Algorithm 1 solves a quadratic programming sub-problem obtained based on the approximation in each iteration. Considering comprehensively, the calculation process of Algorithm 1 is mainly divided into three stages: First of all, update the Hessian matrix of the Lagrangian function based on~(\ref{eqBFGS}). Then, solve the quadratic programming sub-problem, and perform a constraint one-dimensional search, thus we can obtain the maximum network capacity value in the end. Therefore, the complexity of each iteration of Algorithm 1 mainly comes from updating $\mathbf{H}_{k}$ based on $\mathbf{H}_{k-1}$ by~(\ref{eqBFGS}), solving a quadratic programming sub-problem, and performing the constraint one-dimensional search. To note, the BFGS correction method yields the complexity of ${\rm O}(M)$ with the output of $\mathbf{H}_{k}$. On the other hand, the method of using the BFGS correction formula to approximate the Hessian matrix of the objective function for solving a quadratic programming sub-problem can be regarded as a kind of quasi- Newton method. Considering the super-linear convergence of the quasi-Newton method, the complexity of solving the quadratic programming sub-problem is ${\rm O}(M\ln \ln (\frac{1}{{{\varepsilon _1}}}))$, where the termination criteria $\varepsilon_1 $ determines the accuracy of solving the quadratic programming sub-problem. Finally the bisection method is applied with the complexity of ${\rm O}({M^2}{\log _2}(\frac{1}{{{\varepsilon _2}}}))$~\cite{ConOpt}, where the termination criteria $\varepsilon_2 $ determines the accuracy of performing the constraint one-dimensional search. Therefore, Algorithm 1 yields the per-iteration complexity of ${\rm O}(M + M\ln \ln (\frac{1}{\varepsilon_1 }) + M^2{\log _2}(\frac{1}{\varepsilon_2 }))$, which is of polynomial-time computational complexity.

\begin{algorithm}[!t]
\caption{Quadratic Programming Algorithm based on Lagrangian Function in HSR Scenarios}
\label{alg1}
\begin{algorithmic}[1]
\STATE 
Determine
the locations of the
BS and MRs;
\STATE
Locate all users in this train-ground communication system and establishes the relationship between each user and the specific server;
\STATE Construct global network capacity function according to obtained communication system parameters;
\STATE Construct the Lagrangian function of the bandwidth allocation problem based on the objective function and constraints; 
\STATE Determine the iteration starting point ${\boldsymbol{\alpha}_0}$ and the convergence accuracy $\sigma$;
\STATE Approximate the original bandwidth allocation optimization problem to a new quadratic programming sub-problem at point ${\boldsymbol{\alpha}_0}$;
\STATE Solve the obtained quadratic programming sub-problem, and let ${\mathbf{S}_{\rm{0}}} = \mathbf{S}^{\rm{*}}$;
\STATE Perform a constrained one-dimensional search on the global network capacity function in the direction ${\mathbf{S}_{\rm{0}}}$ to obtain a new bandwidth allocation factor ${\boldsymbol{\alpha}_{\rm{1}}}$;
\STATE $k=0$;
\WHILE{${\boldsymbol{\alpha}_{k+1}}$ does not meet the convergence accuracy}
\STATE $k = k + 1$;
\STATE Modify matrix ${\mathbf{H}_k}$
based on
${\mathbf{H}_{k-1}}$;
\STATE Simplify the original bandwidth allocation optimization problem into a new quadratic programming sub-problem at point ${\boldsymbol{\alpha}_k}$;
\STATE Solve the obtained quadratic programming sub-problem and let ${\mathbf{S}_k}=\mathbf{S}^{\rm{*}}$;
\STATE Perform a constrained one-dimensional search on the objective function of the original bandwidth allocation optimization problem in the direction ${\mathbf{S}_k}$, and then obtain the new bandwidth allocation factor ${\boldsymbol{\alpha}_{k + 1}}$;
\ENDWHILE
\STATE Output optimal bandwidth allocation factor ${\boldsymbol{\alpha}^{\rm{*}}}{\rm{ = }}{\boldsymbol{\alpha}_{k + 1}}$;
\STATE Output Maximum network capacity value $f^* = f({\mathbf{X}_{k + 1}})$;
\end{algorithmic}
\end{algorithm}

\section{Performance Evaluation}\label{S5}

\subsection{Simulation Setup}\label{S5-1}

Based on the system model
described
in Section~\ref{S3-1},
the
entire
train-ground communication system works in the 60GHz mm-wave frequency band. On the other hand, because mm-wave
BSs
are commonly used in small-scale and dense deployment scenarios, the simulation environment is set
to be
a square area with
500$\times$500 m$^2$. 200 users randomly appear in this area, and their position coordinates meet the uniform distribution of this area, and each user is associated with the nearest service device, i.e., the principle of ``who is near, who serves." The track-side BS is located above the center of the area. Considering the actual operation scenario of the HSR, the 9 MRs are arranged horizontally below the center of the area. The
transmit
power of the BS and
MRs are $P_{t}$.

For mm-wave communications,
we use the widely used real directional antenna model from IEEE 802.15.3c~\cite{80215}. The model includes a linearly scaled Gaussian main lobe and a constant-level side lobe. Based on this model, the gain of a directional antenna in units of dB can be expressed as
\begin{align}
G({\theta})=\left\{ \hspace{-0.05in}
\begin{array}{ll}
G_{0}-3.01\cdot\left({\frac{2\theta}{\theta_{-3dB}}}\right)^2,& 0^{\circ} \leq \theta \leq \frac{\theta_{ml}}{2} \\
G_{sl}, & \frac{\theta_{ml}}{2} \leq \theta \leq 180^{\circ},\\
\end{array} \right. \label{eq3-4}
\end{align}
where $\theta \in [{0^\circ },{180^\circ }]$, ${G_0}$ is the maximum antenna gain, which
is given by
${G_0} = 10 {\log_{10}}{\left( \frac{{1.6162}}{{\sin \left( {{\theta_{-3dB}}}/2 \right)}} \right)^2}$,
${\theta _{ - 3dB}}$ represents the half-power beamwidth, and ${\theta _{ml}}$ is the main lobe
beamwidth
in degrees.
The relationship between ${\theta _{ml}}$ and ${\theta _{ - 3dB}}$ can be expressed as ${\theta _{ml}} = 2.6 \cdot {\theta _{ - 3dB}}$. The sidelobe gain ${G_{sl}}$
is given by ${G_{sl}} =  - 0.4111 \cdot \ln ({\theta _{ - 3dB}}) - 10.579$.
Other simulations parameters are given in Table~\ref{table1}.

\begin{table}[!t]
\begin{center}
\caption{Simulation Parameters}
\begin{tabular}{lcc}
\toprule
Parameter & Symbol  & Value \\
\midrule
Transmit power & $P_{t}$ & 1000 mW  \\
Path loss exponent  &  $n$   &  2 \\
Transceiver efficiency factor & $\eta $ & 0.5 \\
Background noise &  $N_{0}$  & -134 dBm/MHz \\
Half-power beamwidth & $\theta_{-3dB}$ & $30^{\circ}$ \\
Link blockage probability on average & $P_{b}$ & $0.2$ \\
\bottomrule
\end{tabular}
\label{table1}
\end{center}
\end{table}

In order to evaluate the performance of the
proposed
SQP algorithm,
the following four algorithms
are chosen as baseline schemes:
\begin{enumerate}
\item PNOU (Priority based on the number of users): Considering the established communication scenario, the location of each user meets a random and uniform distribution in the
communication system area. According to the ``shortest distance criterion,"
when the association relationship between the user and the track-side BS or the MR is determined in advance, the BS and MRs can count the number of users associated with themselves, and use the number of associated users as the criterion for how much bandwidth is allocated. BS or MRs with a larger number of associated users will be allocated
with
relatively more bandwidth resources.

\item PD (Priority based on distance):
After determining the association relationship between each user and the BS or MR according to the ``shortest distance criterion,"
the BS and the MRs can collect the location information of their respective associated users,
save
the
distance from each associated user, and use the average associated users'
distance as the criterion for how much bandwidth is allocated. BS or MRs with a smaller average associated users' distance will be allocated relatively more bandwidth resources.

\item IP (Interior point algorithm):
The interior point algorithm
transforms
the original
constrained
optimization problem
into
an
unconstrained optimization problem.
When constructing the objective function of
the
equivalent unconstrained optimization problem,
it
defines the objective function as a penalty function in the feasible region.
It
then solves the extreme point of penalty function in the feasible region.
That is,
the interior point algorithm ensures that the iteration point selected when solving the equivalent unconstrained optimization problem is always within the feasible region,
as indicated by
``interior point." When the selected iteration point moves from the feasible region to the constrained boundary, the value of the penalty function will increase sharply to infinity, which acts as a penalty to ensure that the iterative point will never touch the constrained boundary, so as to continuously approach the optimal solution of the original constrained optimization problem within the feasible region.

\item TR (Trust region algorithm):
The fundamental idea of the trust region (TR) algorithm is a little different from that of the SQP algorithm. The SQP algorithm follows the principle of ``adjusting the search direction and performing one-dimensional search."
The TR algorithm,
on the other hand,
hopes to find the optimal solution of the problem in a domain. The ``trust region" generally refers to a small neighborhood of the current iteration point. In this neighborhood, a sub-problem of the original nonlinear programming problem is solved to obtain the trial step size ${S^k}$. Then 
an
evaluation function is used to determine whether to accept the trial step size and to determine the trust region
within
which the next round of iterative calculation
will be done.
At this time, if accept, then ${\mathbf{X}_{k + 1}} = {\mathbf{X}_k} + {\mathbf{S}_k}$; otherwise ${\mathbf{X}_{k + 1}} = {\mathbf{X}_k}$. The size of the new trust region is determined according to the quality of the trial step. If the trial step is good, the trust region will expand or remain unchanged in the next round of iterative calculation; otherwise, the trust region will decrease. It is worth noting that some key mathematical steps of the TR algorithm are consistent with the SQP algorithm, but the fundamental ideas of the two are different. In many iterative calculations, the TR algorithm keeps updating the search radius, while the SQP algorithm keeps updating the search direction.
\end{enumerate}

Among the four comparison schemes listed above, the first two are greedy algorithms, i.e., they do not consider the overall optimality and always use a macro strategy based on a certain fixed idea, which is easier to implement. The latter two are common optimization algorithms for solving nonlinear programming problems. Their fundamental ideas have certain similarities with that of the SQP algorithm. They are
also
gradually approaching the optimal solution through
iterations, but
follow different specific guidelines in the search.

According to the optimization goal of
problem (P1),
the network capacity of the entire train-ground communication system, i.e., the sum of the average user's transmission rate of
the
BS and the average users' transmission rate of all MRs is used as performance metric, which is the optimization goal of this research.
\begin{align}
\max &~~\bar{R}_{BS} + \sum_{i = 1}^{Rnum} {\bar{R}_i} \label{goal}
\end{align}

We will simulate the
performance of the algorithms
under different
total
bandwidth $W$ and different FD SI cancellation levels, i.e., $\beta$, at the MRs.


\subsection{Simulation Results and Discussions}\label{S5-2}


\subsubsection{Under Different
	Total Bandwidth $W$}\label{S5-2-1}
\begin{figure}[!t]
\begin{center}
\includegraphics*[width=1\columnwidth,height=2.6in]{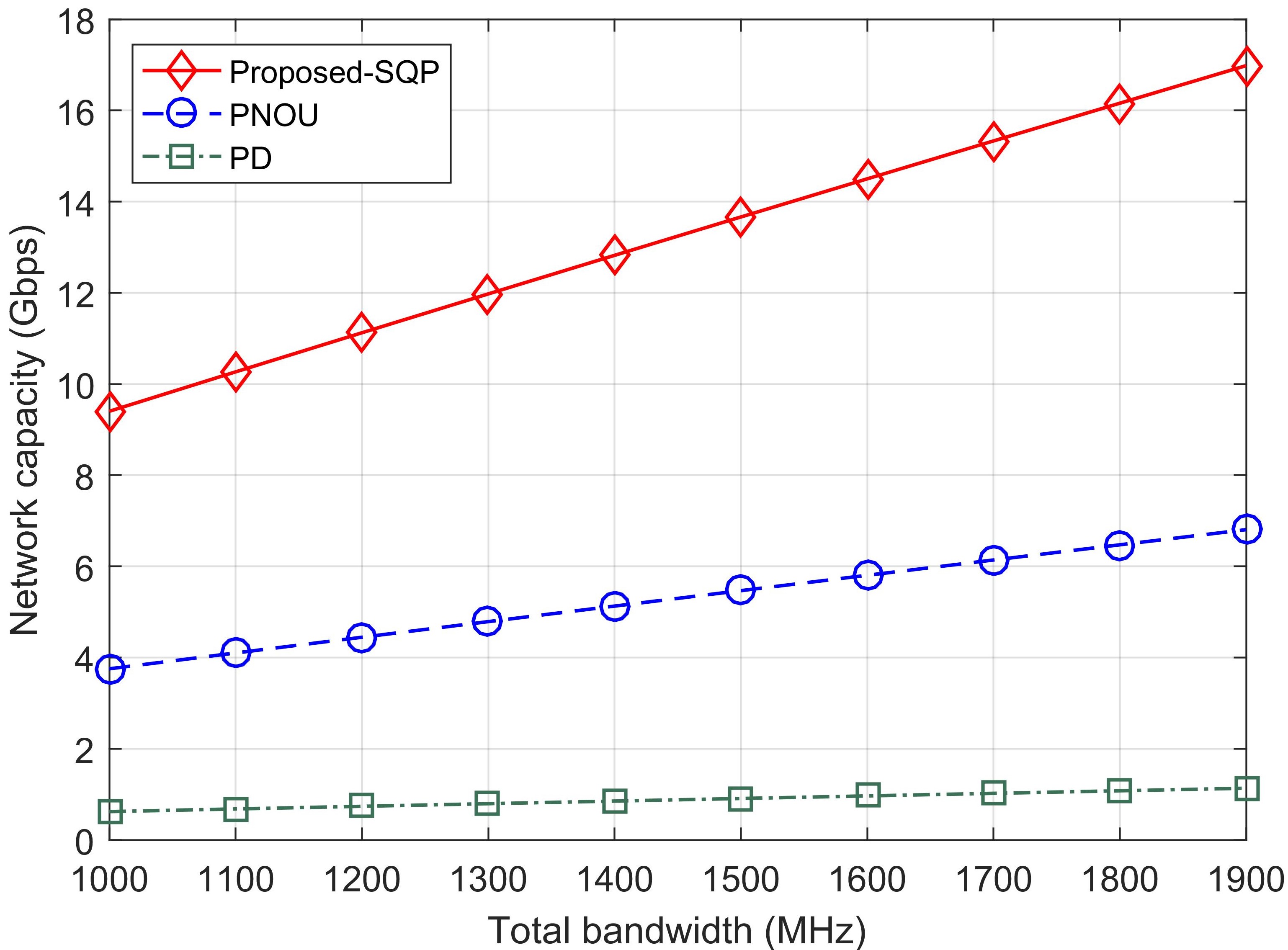}
\end{center}
\caption{Network capacity under different
	total bandwidth $W$.} \label{fig4-6}
\end{figure}

In this simulation, we apply
the two greedy algorithms and the
proposed
SQP algorithm
to evaluate
the changes in network capacity as the
system bandwidth $W$ is
increased.
The simulation results are
shown in Fig.~\ref{fig4-6}.
For more rigorous, we display the data obtained by this simulation in Table~\ref{tablefig1}.
In this simulation,
the SI cancellation capability of all FD MRs is
set to $10^{-7}$.

\begin{table}[!t]
	\begin{center}
		\caption{Network capacity under different total bandwidth $W$}
		\begin{tabular}{c|c|c|c}
			\toprule
			Total bandwidth (MHz) & \multicolumn{3}{c}{Network capacity (Gbps)} \\
			\midrule
			~ & Proposed-SQP & PNOU & PD \\
			\midrule
			1000 & 9.401 & 3.753 & 0.621 \\
			1100 & 10.266 & 4.101 & 0.679 \\
			1200 & 11.124 & 4.445 & 0.737 \\
			1300 & 11.976 & 4.788 & 0.794 \\
			1400 & 12.822 & 5.128 & 0.851 \\
			1500 & 13.663 & 5.467 & 0.908 \\
			1600 & 14.501 & 5.804 & 0.965 \\
			1700 & 15.331 & 6.139 & 1.021 \\
			1800 & 16.159 & 6.473 & 1.077 \\
			1900 & 16.983 & 6.805 & 1.133 \\
			\bottomrule
		\end{tabular}
		\label{tablefig1}
	\end{center}
\end{table}

It can be seen from
Fig.~\ref{fig4-6} that
as the total amount of bandwidth is increased,
the three algorithms all
exhibit a trend of
linear growth.
The performance of the
proposed
SQP algorithm is significantly better than
that of
PNOU and PD.
Furthermore, as the
system bandwidth
is increased,
the network capacity achieved by the SQP algorithm
increases at a faster rate than the two baseline schemes.
%
With the proposed
SQP algorithm, the sensitivity of the network capacity
to the
system bandwidth
reaches about 9Mbps/100MHz, which is significantly higher than the 4.25Mbps/100MHz
achieved by
the
PNOU.
With PD, the network capacity value
does
not increase significantly
even
as the total bandwidth is increased by 900MHz.
Especially, when the
total bandwidth is 1200MHz, the network capacity achieved by the SQP algorithm
is
about 11.2Gbps, which is about
2.6 times
of  4.4Gbps achieved by the PNOU, and about 12.4 times
of 0.9Gbps achieved by the PD.
When the total bandwidth is increased from 1200MHz to 1700MHz, the network capacity achieved by the SQP algorithm increases by about 4.1Gbps, which is about 2.4 times of 1.7Gbps achieved by the PNOU, and about 20.5 times of 0.2Gbps achieved by the PD.

As mentioned
above,
the user information collected by the global train-ground communication network is only
each user's location and
the
respective distances to the server.
The network performance
of
PD is the worst. This is because the BS and MRs are not evenly distributed in the established HSR scenario. According to the users' positions,
except for some regional edge users,
the
BS
is
associated with users above the entire area. Therefore, users associated with
the
BS are more dispersed than that associated with
the
MRs. On the other hand, it is obvious that there are more users associated with
the
BS, and the PD algorithm
allocates less bandwidth to the BS. This is
why the average users' transmission rate of
the
BS cannot be
effectively
improved with PD.

\begin{figure}[!t]
\begin{center}
\includegraphics*[width=1\columnwidth,height=2.6in]{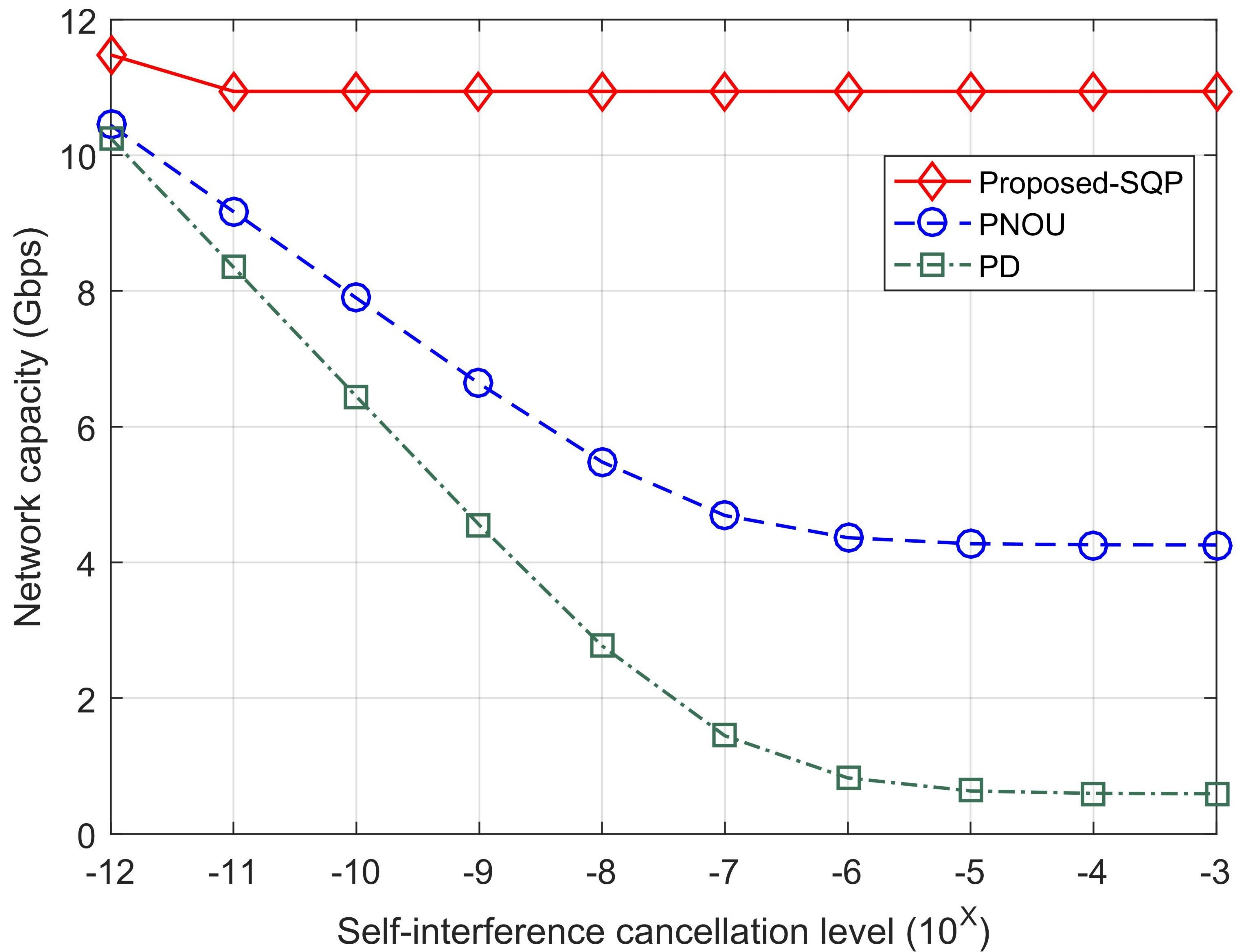}
\end{center}
\caption{Network capacity under different levels of SI cancellation.} \label{fig4-7}
\end{figure}

\begin{table}[!t]
	\begin{center}
		\caption{Network capacity under different levels of SI cancellation}
		\begin{tabular}{c|c|c|c}
			\toprule
			$\beta$ & \multicolumn{3}{c}{Network capacity (Gbps)} \\
			\midrule
			~ & Proposed-SQP & PNOU & PD \\
			\midrule
			$1.0{\rm{ \times }}10^{-12}$ & 11.477 & 10.441 & 10.251\\
			$1.0{\rm{ \times }}10^{-11}$& 10.943 & 9.168 & 8.349 \\
			$1.0{\rm{ \times }}10^{-10}$ & 10.943 & 7.894 & 6.444 \\
			$1.0{\rm{ \times }}10^{-9}$ & 10.943 & 6.635 & 4.552 \\
			$1.0{\rm{ \times }}10^{-8}$ & 10.942 & 5.479 & 2.771 \\
			$1.0{\rm{ \times }}10^{-7}$ & 10.942 & 4.691 & 1.444 \\
			$1.0{\rm{ \times }}10^{-6}$ & 10.941 & 4.363 & 0.823 \\
			$1.0{\rm{ \times }}10^{-5}$& 10.941 & 4.274 & 0.631 \\
			$1.0{\rm{ \times }}10^{-4}$ & 10.941 & 4.259 & 0.595 \\
			$1.0{\rm{ \times }}10^{-3}$ & 10.941 & 4.257 & 0.591 \\
			\bottomrule
		\end{tabular}
		\label{tablefig2}
	\end{center}
\end{table}

\subsubsection{Under Different Levels of SI Cancellation $\beta$ }\label{S5-2-2}

Since the MRs adopt the FD communication mode,
they also suffer residual SI. In order to study the performance of different algorithms in response to a sharp increase of SI, we also
simulate
the network capacity
under different levels of SI cancellation.
The results are presented in Fig.~\ref{fig4-7}, and the specific experimental data are shown in Table~\ref{tablefig2}.
In this simulation,
the initial system total bandwidth is
fixed at
1200MHz.

It can be seen from Fig.~\ref{fig4-7} that as the SI cancellation capability of
the
MRs
is weakened,
the network capacity
achieved by
the PNOU and PD both
exhibit
an
exponential decline. When
$\beta$ is increased above
${10^{-5}}$, the network capacity
achieved by the PNOU and
PD reaches the lowest value, which is about 4.2Gbps and 0.5Gbps, respectively.
Then they no longer decrease
as $\beta$ is further increased.

In contrast,
the network capacity
achieved
by the SQP algorithm
remains at a relatively high level, as $\beta$ is increased.
Especially, when $\beta$ is increased to ${10^{-5}}$, the network capacity achieved by the SQP is about 11Gbps, which is significantly better than 4.2Gbps of the PNOU and 0.5Gbps of the PD.
When $\beta$ is increased from ${10^{-12}}$ to ${10^{-5}}$, the network capacity achieved by the SQP algorithm decreases by about 0.5Gbps, which is about 0.08 times of 6.3Gbps achieved by the PNOU, and about 0.05 times of 9.7Gbps achieved by the PD.
In other words, the
proposed
SQP algorithm has the ability to perceive the sudden increase of SI, and can coordinate the bandwidth resources of the entire network to ensure a consistently high
network capacity.
Considering that in the simulation, the SI cancellation level of each MR is the same,
when the SI cancellation level drops significantly,
the performance of the entire FD MR communication system drops sharply. We initially guessed that in order to ensure the network capacity of train-ground communication system, the
SQP algorithm
would make
a decision to allocate almost all
the
bandwidth resources to
the
track-side BS operating in
the HD
mode. In addition, the number of users associated with the BS is relatively large, so the
SQP algorithm
still guarantees a high network capacity value
under sharply increased $\beta$.

To
verify this conjecture,
the optimal bandwidth allocation factor ${\boldsymbol{\alpha}}$ computed
by the SQP algorithm when the SI cancellation level is $10^{-3}$
is
shown in Table~\ref{table3}. It can be found that when the SI cancellation level of the MRs falls on the extremely poor order of magnitude, the bandwidth allocation decision obtained by the
SQP algorithm is that the BS
is assigned with
almost all
the
bandwidth resources, which
confirms the previous conjecture.


\begin{table}[!t]
\begin{center}
\caption{The ${\boldsymbol{\alpha}}$ Obtained by SQP When the SI Cancellation Level is $10^{-3}$}
\begin{tabular}{c|c|c}
\toprule
${{\bf{\alpha }}_i}$ & Associated equipment & Value \\
\midrule
${{\bf{\alpha }}_0}$ & BS & 1.0000 \\
${{\bf{\alpha }}_1}$ & MR1 & 1.7598e-09  \\
${{\bf{\alpha }}_2}$ & MR2 & 3.9526e-09 \\
${{\bf{\alpha }}_3}$ & MR3 & 1.1182e-09 \\
${{\bf{\alpha }}_4}$ & MR4 & 4.7802e-09 \\
${{\bf{\alpha }}_5}$ & MR5 & 2.2247e-09 \\
${{\bf{\alpha }}_6}$ & MR6 & 5.1305e-09 \\
${{\bf{\alpha }}_7}$ & MR7 & 1.8028e-09 \\
${{\bf{\alpha }}_8}$ & MR8 & 1.2927e-09 \\
${{\bf{\alpha }}_9}$ & MR9 & 1.3782e-09 \\
\bottomrule
\end{tabular}
\label{table3}
\end{center}
\end{table}

On the other hand,
once the location of each user is determined in the train-ground communication system, the association relationship between
a
user
is determined, and the
bandwidth allocation factors obtained by
PNOU and
PD
are
determined. When the SI cancellation capability of the FD MRs
is
too weak,
their received signals
will
be overwhelmed by the SI, and the user's transmission rate
will be
almost zero.
The contribution of the average user's transmission rate
through
the MRs to the overall network capacity will be
almost zero.
This is the
reason why the network capacity
reaches its lowest
under PNOU and PD
when the SI cancellation capability of the MRs is too small.
The lowest capacity
mainly comes from
the HD BS, because it is not limited
by
the SI cancellation capability of the MRs. When the level of SI cancellation changes, as long as the bandwidth resources allocated to the BS remain unchanged,
the users' transmission rates associated with the BS will not be affected.

With the above considerations and reviewing Fig.~\ref{fig4-7}, we
find that when the SI cancellation level at the FD MRs
is
between
$10^{-11}$ and $10^{-3}$, the global network capacity is basically stable at about 11Gbps.
Does
this mean that,
when the SI cancellation level at the FD MRs drops
to
$10^{-11}$, the SQP algorithm directly makes the decision shown in Table~\ref{table3}, i.e.,
almost all
the
bandwidth resources are straightforwardly allocated to the BS?
We run two more sets
of simulations, which are also based on the simulation parameters shown in Table~\ref{table1}. The difference between the two sets of simulations is that, the first set of simulations are carried out
for $\beta =  10^{-9}$, and
the second set is for $\beta =  10^{-6}$.
The bandwidth allocation results achieved by the SQP algorihtm are presented in
Table~\ref{table4}.


\begin{table}[!t]
	\begin{center}
		\caption{The ${\boldsymbol{\alpha}}$ Obtained by SQP When the SI Cancellation Level is $10^{-9}$ and $10^{-6}$}
		\begin{tabular}{c|c|c|c}
			\toprule
			${{\bf{\alpha }}_i}$ & Associated equipment & Value & Value \\
			\midrule
			& & $\beta=10^{-9}$ & $\beta=10^{-6}$ \\
			\midrule
			${{\bf{\alpha }}_0}$ & BS & 0.1921 & 0.3952 \\
			${{\bf{\alpha }}_1}$ & MR1 & 0.0836 & 0.0161 \\
			${{\bf{\alpha }}_2}$ & MR2 & 0.0344 & 0.0196 \\
			${{\bf{\alpha }}_3}$ & MR3 & 0.0527 & 0.1159 \\
			${{\bf{\alpha }}_4}$ & MR4 & 0.1504 & 0.1383 \\
			${{\bf{\alpha }}_5}$ & MR5 & 0.0825 & 0.0306 \\
			${{\bf{\alpha }}_6}$ & MR6 & 0.0881 & 0.0132 \\
			${{\bf{\alpha }}_7}$ & MR7 & 0.1440 & 0.0146 \\
			${{\bf{\alpha }}_8}$ & MR8 & 0.0899 & 0.1259 \\
			${{\bf{\alpha }}_9}$ & MR9 & 0.0822 & 0.1306 \\
			\bottomrule
		\end{tabular}
		\label{table4}
	\end{center}
\end{table}

From Table~\ref{table4},
it can be seen
that when the SI cancellation level of the FD MRs deteriorates, the
SQP algorithm does not allocate nearly all the bandwidth resources
to the track-side BS.
As the level of SI cancellation deteriorates, more bandwidth resources are gradually allocated to
the
BS. This decision is in line with the actual situation, i.e., although the level of SI cancellation at the MRs
has
deteriorated, if the SI has not reached the level of overwhelming desired signals, the normal communications
of users associated with the MR should
still be
helpful.
This helps maintain fairness between users associated with the
BS and MRs. From this point of view, in Fig.~\ref{fig4-7}, the network capacity achieved by the SQP algorithm
is stabilized
at 11Gbps.

\subsubsection{Comparison with Other Optimization Algorithms}\label{S5-2-3}
\begin{figure}[!t]
\begin{center}
\includegraphics*[width=1\columnwidth,height=2.7in]{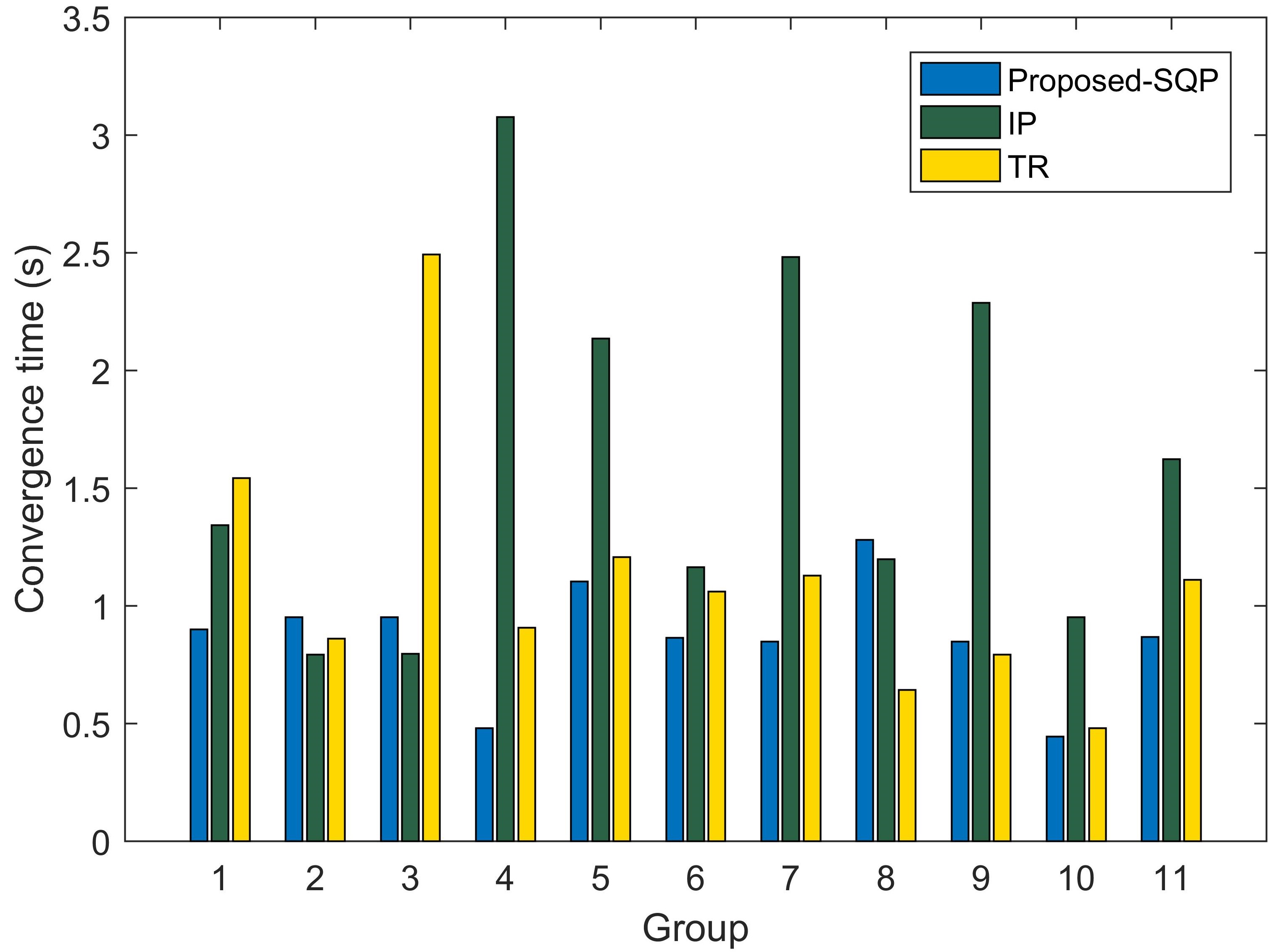}
\end{center}
\caption{Convergence time
	of the
	three optimization algorithms.} \label{figtable4-3-1}
\end{figure}

\begin{figure}[!t]
\begin{center}
\includegraphics*[width=1\columnwidth,height=2.7in]{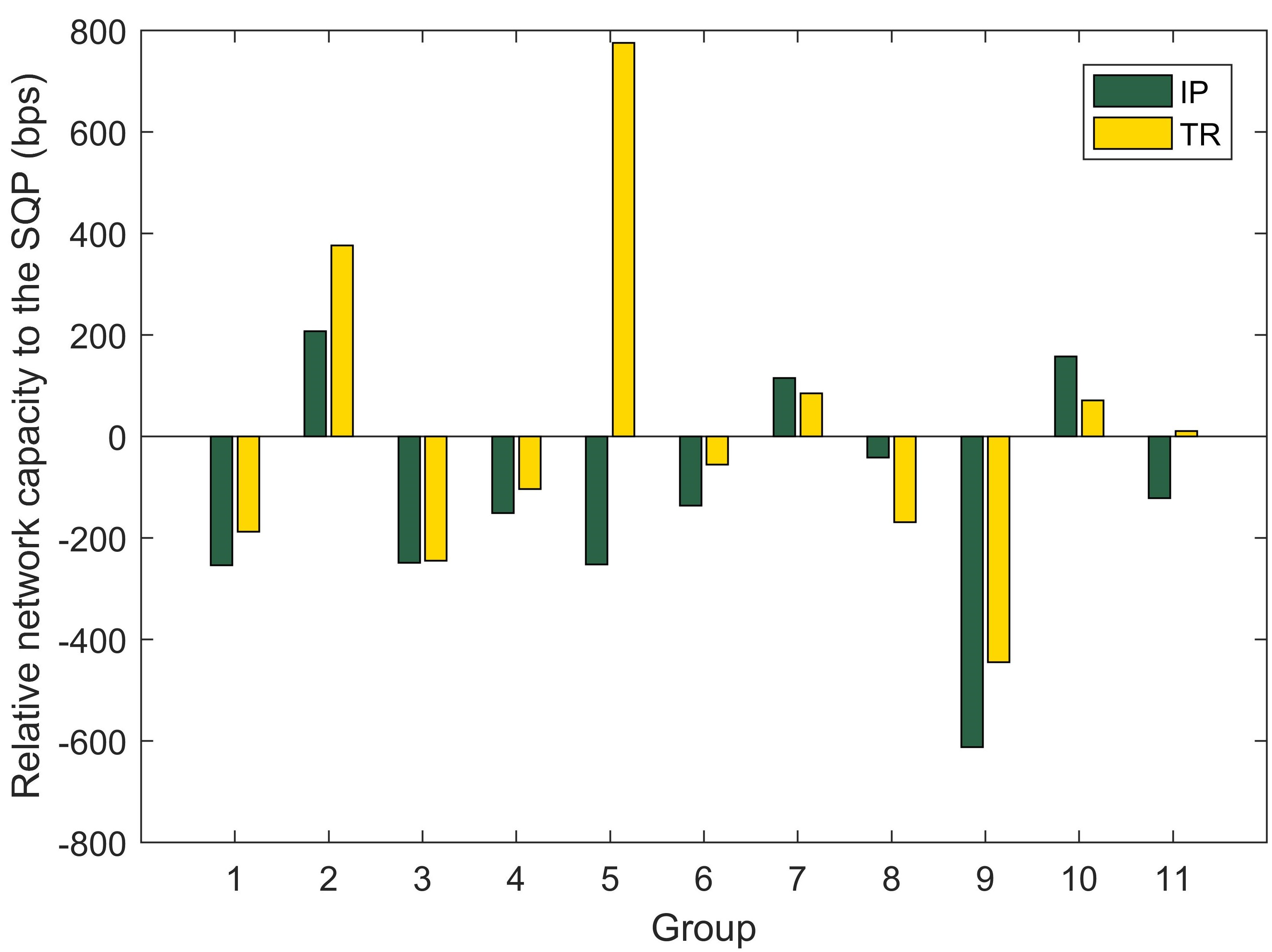}
\end{center}
\caption{Relative value of network capacity
	achieved by the
	three optimization algorithms.} \label{figtable4-3-2}
\end{figure}

We also used two other
optimization algorithms
for constrained nonlinear programming problems,
i.e., IP and TR, as baseline schemes.
In the simulations, we
find that if other simulation parameters are the same, there is only a difference of less than 1Kbps between the network capacity
achieved
by the
proposed
SQP algorithm and that achieved by
the IP and the TR algorithm.
Since the network capacity of the train-ground communication system
is on the order of Gbps,
such a small difference is negligible.
Therefore,
network capacity under different
system bandwidths and different SI cancellation levels obtained by the
proposed SQP algorithm are
close to that achieved by the IP and the TR algorithm.
Instead, we compare the three optimization schemes with respect to
their execution time
and the relative value of the maximum network capacity obtained. The results are
presented in Figs.~\ref{figtable4-3-1} and~\ref{figtable4-3-2},
where
the 11th groups
of data represent the average of the previous 10 groups of data, and and the specific experimental data are shown in Table~\ref{tablefig3} and Table~\ref{tablefig4}.
Note
that the simulation parameters of each group in Figs.~\ref{figtable4-3-1} and~\ref{figtable4-3-2}
are given in
Table~\ref{table1};
the
total
bandwidth resource
is fixed at
1200MHz; and
the SI cancellation level
of the MRs is
fixed at
$10^{-7}$.
In
the beginning of each group of experiments,
the location and associated information of 200 users
are
randomly generated,
and finally the average value of
the
10 groups of
experiments
will be calculated and presented.

\begin{table}[!t]
	\begin{center}
		\caption{Convergence time of the three optimization algorithms}
		\begin{tabular}{c|c|c|c}
			\toprule
			Group & \multicolumn{3}{c}{Convergence time (s)} \\
			\midrule
			~ & Proposed-SQP & IP & TR \\
			\midrule
			1 & 0.901 & 1.343 & 1.542 \\
			2 & 0.952 & 0.792 & 0.861 \\
			3 & 0.952 & 0.796 & 2.492 \\
			4 & 0.481 & 3.076 & 0.908 \\
			5 & 1.104 & 2.136 & 1.208 \\
			6 & 0.864 & 1.164 & 1.061 \\
			7 & 0.848 & 2.248 & 1.128 \\
			8 & 1.281 & 1.199 & 0.643 \\
			9 & 0.848 & 2.288 & 0.792 \\
			10 & 0.444 & 0.952 & 0.481 \\
            11 & 0.867 & 1.623 & 1.111 \\
			\bottomrule
		\end{tabular}
		\label{tablefig3}
	\end{center}
\end{table}

\begin{table}[!t]
	\begin{center}
		\caption{Relative value of network capacity achieved by the three optimization
algorithms}
		\begin{tabular}{c|c|c}
			\toprule
			Group & \multicolumn{2}{c}{Relative network capacity to the SQP (bps)} \\
			\midrule
			~ & IP & TR \\
			\midrule
			1 & ~~~~~~~254.15~~~~~~~ & 187.65 \\
			2 & ~~~~~~~-207.34~~~~~~~ & -376.03 \\
			3 & ~~~~~~~248.94~~~~~~~ & 245.2 \\
			4 & ~~~~~~~151.03~~~~~~~ & 103.74 \\
			5 & ~~~~~~~251.84~~~~~~~ & -775.57 \\
			6 & ~~~~~~~136.11~~~~~~~ & 55.243 \\
			7 & ~~~~~~~-115.49~~~~~~~ & -85.262 \\
			8 & ~~~~~~~41.915~~~~~~~ & 168.69 \\
			9 & ~~~~~~~612.18~~~~~~~ & 444.62 \\
			10 & ~~~~~~~-157.51~~~~~~~ & -71.381 \\
            11 & ~~~~~~~121.58~~~~~~~ & -10.309 \\
			\bottomrule
		\end{tabular}
		\label{tablefig4}
	\end{center}
\end{table}


It should be noted that when using any iterative optimization algorithm to solve
a
nonlinear programming problem, it is necessary to set the parameter
of accuracy requirement, and the iteration
process
stops when the result that satisfies the
accuracy
requirement
is obtained. Therefore, for a specific optimization algorithm, increasing the
required
accuracy will inevitably lead to an increase in
execution
time. In the above 10 sets of experiments, the accuracy
requirements
of the three algorithms are set to the Kbps level in advance. This allows us to compare the convergence speed of the three algorithms.

First observe the average
convergence time shown
in Fig.~\ref{figtable4-3-1}. From the point of view of
execution time,
the proposed
SQP algorithm
outperforms both IP and TR, and TR is faster than IP.
Also the convergence time of the proposed SQP algorithm is the most stable one among the three algorithms.
In Fig.~\ref{figtable4-3-2}, with the simulation parameters and convergence accuracy fixed, taking the data of the fifth group as an example, the maximum network capacity value calculated by the SQP algorithm is 0.252Kbps higher than that calculated by IP, and 0.776Kbps lower than that calculated by TR. From the calculated average relative value of the maximum network capacity shown in Fig.~\ref{figtable4-3-2}, the maximum network capacity value obtained by the
SQP algorithm is
0.122Kbps
lower
than
that
obtained by
IP,
and 0.01Kbps larger than
that
obtained by TR.
The IP algorithm has the longest average running time,
but also with a slightly
better network capacity.
It may not worth pursuing for such a small gain in capacity at the greatly extended execution time.
The average running time of the TR algorithm is in the middle of the three algorithms, but the average network capacity value obtained by TR is lower than that of the SQP algorithm.
On the other hand, fewer stations are set along the high-speed railways, thus ensuring that the high-speed trains can maintain a long-term high-speed and stable operation. When the train is running at a high and stable speed, the moving line is deterministic and the train's position is predictable, so the power of the received signal is also highly predictable. This feature makes it easier to optimize the management of the mobility and resources of the communication system in HSR scenarios. In the investigated mm-wave train-ground communication system, if the train's position can be predicted, the relative distance between MRs and users can be determined, and the bandwidth allocation optimization problem can be formulated for a period of time in the future, and the proposed algorithm can also make decisions in advance.

In the above simulations, we also
find
an interesting phenomenon: When the simulation parameters are unchanged and the
SQP algorithm is run multiple times to solve the P1, the maximum network capacity value obtained within the convergence accuracy corresponds to a variety of optimal bandwidth allocation factors. The values in these optimal bandwidth allocation factors are slightly different, but the maximum network capacity values obtained by using them are the same. This also explains that
problem (P1) has multiple local optimal solutions, and there are several local optimal solutions that are within the convergence accuracy of the global optimal solution, so they can all be directly regarded as the global optimal solution. From the basic idea of the
SQP algorithm, this means that there are many search directions for the global optimal solution, which greatly accelerates the convergence speed, reduces the number of algorithm iterations, and improves the
performance of the
proposed
SQP algorithm.

\section{Conclusions}\label{S6}

In this paper, we introduced
a SQP algorithm
based on
the
Lagrangian function
to solve the problem of bandwidth allocation between the track-side BS and FD MRs in
a train-ground communication system. The SQP algorithm ensures that the optimal solution of the problem can be quickly approached, and the Lagrangian function ensures that the sub-problems of the original problem approximated
in
each iteration are
convex,
i.e., the local optimal solution of the problem is the global optimal solution. Extensive simulation results demonstrated that the
proposed
SQP algorithm can effectively improve the network capacity of the train-ground communication in HSR scenarios while maintaining a certain anti-SI ability,
when compared with four baseline schemes.
For the future work, we will deeply investigate the blockage problem in the studied mm-wave train-ground communication system, and consider using IRS to propose a solution.


\begin{thebibliography}{10}

\bibitem{Ai}
B. Ai, A. F. Molisch, M. Rupp and Z. Zhong, ``5G key technologies for smart railways," in Proceedings of the IEEE, vol. 108, no. 6, pp. 856-893, June 2020.

\bibitem{Doppler}
D. Fan, Z. Zhong, G. Wang, and F. Gao, ``Doppler shift estimation for high-speed railway wireless communication systems with large-scale linear antennas," in Proc. Int. Workshop High Mobility Wireless Commun., pp. 96-100, Oct. 2015.

\bibitem{HSRMR}
M. Gao et al., ``Dynamic mmWave beam tracking for high speed railway communications," in Proc. IEEE Wireless Commun. Netw. Conf. Workshops (WCNCW), pp. 278-283, Apr. 2018.

\bibitem{ROF}
B. Lannoo, D. Colle, M. Pickavet, and P. Demeester, ``Radio-over-fiber-based solution to provide broadband Internet access to train passengers,"
{\em IEEE Communications Magazine}, vol. 45, no. 2, pp. 56-62, Feb. 2007.

\bibitem{mmwave5G}
M.~Elkashlan, T. Q.~Duong, and H. H.~Chen, ``Millimeter-wave communications for 5G: fundamentals: Part I [Guest Editorial]," {\em IEEE Communications Magazine}, vol. 52, no. 9, pp. 52-54, Sept. 2014.

\bibitem{FDintr}
Z. Zhang, X. Chai, K. Long, A. V. Vasilakos and L. Hanzo, ``Full duplex techniques for 5G networks: self-interference cancellation, protocol design, and relay selection," {\em IEEE Communications Magazine}, vol. 53, no. 5, pp. 128-137, May 2015.

\bibitem{Des60GHz}
C. H. Doan, S. Emami, D. A. Sobel, A. M. Niknejad, and R. W. Brodersen, ``Design considerations for 60 GHz CMOS radios,'' {\em IEEE Communications Magazine}, vol. 42, no. 12, pp. 132-140, Dec. 2004.

\bibitem{ECMC}
ECMC TC48, ``ECMA standard 387--High rate 60 GHz PHY, MAC and HDMI PAL,'' 2nd edition Dec. 2010.

\bibitem{80215}
IEEE 80215 WPAN Millimeter Wave Alternative PHY Task Group 3c (TG3c), ``Wireless Medium Access Control (MAC) and Physical Layer (PHY) specifications for high rate Wireless Personal Area Networks (WPANs) (Amendement 2: Millimeter-wave-based Alternative Physical Layer Extension),'' Oct. 2009.

\bibitem{80211}
IEEE 80211ad Standard, ``Wireless LAN Medium Access Control (MAC) and Physical Layer(PHY) Specifications (Amendment 3: Enhancements for Very High Throughput in the 60 GHz Band),'' Dec. 2012.

\bibitem{beamintr}
R. W. Heath, N. Gonz¨¢lez-Prelcic, S. Rangan, W. Roh and A. M. Sayeed, ``An overview of signal processing techniques for millimeter wave MIMO systems," {\em IEEE Journal of Selected Topics in Signal Processing}, vol. 10, no. 3, pp. 436-453, April 2016.

\bibitem{pseudowired}
R. Mudumbai, S. K. Singh, and U. Madhow, ``Medium access control for 60 GHz outdoor mesh networks with highly directional links,'' in {Proc. IEEE INFOCOM 2009 (Mini Conference)}, Rio de Janeiro, Brazil, pp. 2871-2875, Apr. 2009.

\bibitem{FD}
A. Sadeghi, M. Luvisotto, F. Lahouti, S. Vitturi, and M. Zorzi, ``Statistical QoS analysis of full duplex and half duplex heterogeneous cellular networks,'' in {\em Proc. IEEE ICC 2016}, Kuala Lumpur, Malaysia, pp. 1-6, May 2016.

\bibitem{Lag}
D. P. Bertsekas, {\em Constrained Optimization and Lagrange Multiplier Methods}, Cambridge, MA: Academic Press, 1982.

\bibitem{CorTra}
I. K. Son, S. Mao, M. X. Gong, and Y. Li, ``On frame-based scheduling for directional mmWave WPANs,'' in {Proc. IEEE INFOCOM 2012}, Orlando, FL, pp. 2149-2157, Mar. 2012.

\bibitem{CorTra2}
I. K. Son, S. Mao, Y. Li, M. Chen, M.X. Gong, and T.S. Rappaport, ``Frame-based medium access control for 5G wireless networks,'' {\em Springer MONET J.}, vol. 20, no. 6, pp. 763-772, Dec. 2015.

\bibitem{FD}
M.~Jain, J.~I. Choi, T.~Kim, D.~Bharadia, S.~Seth, K.~Srinivasan, P.~Levis, S.~Katti, and P.~Sinha, ``Practical, real-time, full duplex wireless,'' in \emph{Proc. ACM MobiCom 2011}, Las Vegas, NV, pp. 301-312, Sept. 2011.

\bibitem{FDRelay1}
H. Cui, M. Ma, L. Song and B. Jiao, ``Relay selection for two-way full duplex relay networks With amplify-and-forward protocol," {\em IEEE Transactions on Wireless Communications}, vol. 13, no. 7, pp. 3768-3777, July 2014.

\bibitem{FDRelay2}
K. M. Rahman, N. Hakem and B. Barua, ``FD-MIMO relay self-interference cancellation using space projection algorithms," 2017 IEEE 9th Latin-American Conference on Communications (LATINCOM), Guatemala City, pp. 1-5, 2017.

\bibitem{niuconvey}
Y.~Niu, Y.~Li, D.~Jin, L.~Su, A. V. Vasilakos, ``A Survey of millimeter wave communications (mmWave) for 5G: Opportunities and challenges,'' {\em Wireless Networks}, vol. 21, no. 8, pp. 2657-2676, Nov. 2015.

\bibitem{niu2017energy}
Y.~Niu, C.~Gao, Y.~Li, L.~Su, D.~Jin, Y.~Zhu, and D.~O. Wu, ``Energy-efficient scheduling for mmwave backhauling of small cells in heterogeneous cellular networks,'' \emph{IEEE Transactions on Vehicular Technology}, vol.~66, no.~3, pp. 2674--2687, Mar. 2017.

\bibitem{chen2018allocation}
Y.~Chen, B.~Ai, Y.~Niu, K.~Guan, and Z.~Han, ``Resource allocation for device-to-device communications underlaying heterogeneous cellular networks using coalitional games,'' {\em IEEE Transactions on Wireless Communications}, vol. 17, no. 6, pp. 4163--4176, June 2018.

\bibitem{energy2}
L. Wang, B. Ai, Y. Niu, X. Chen, and P. Hui, ``Energy-efficient power control of train-ground mmWave communication for high-speed trains,'' {\em IEEE Transactions on Vehicular Technology}, vol. 68, no. 8, pp. 7704--7714, Aug. 2019.

\bibitem{DingzhuWen}
D.~Z. Wen and G.~D. Yu, ``Time-division cellular networks with full-duplex base stations,'' \emph{IEEE Communications Letters}, vol.~20, no.~2, pp. 392--395, Feb. 2016.

\bibitem{dingweiguang}
W.~G. Ding, Y.~Niu, H.~Wu, Y.~Li, and Z.~D. Zhong, ``QoS-aware full-duplex concurrent scheduling for millimeter wave wireless backhaul networks,'' \emph{IEEE Access J.}, vol.~6, pp. 25313--25322, May 2018.

\bibitem{Skouroumounis}
C.~Skouroumounis, C.~Psomas, and I.~Krikidis, ``Heterogeneous FD-mm-Wave cellular networks with cell center/edge users,'' \emph{IEEE Transactions on Communications}, vol.~67, no.~1, pp. 791--806, Jan. 2019.

\bibitem{YibingWang}
Y. Wang, Y. Niu, H. Wu, Z. Han, B. Ai, and Q. Wang, ``Sub-channel allocation for device-to-device underlaying full-duplex mmWave small cells using coalition formation games," {\em IEEE Transactions on Vehicular Technology}, vol. 68, no. 12, pp. 11915--11927, Dec. 2019.

\bibitem{OptimalXiao}
S. Xiao, X. Zhou, Y. Yuan-Wu, G. Y. Li, and W. Guo, ``Robust resource allocation in full-duplex-enabled OFDMA femtocell networks,'' {\em IEEE Transactions on Wireless Communications}, vol. 16, no. 10, pp. 6382--6394, Oct. 2017.

\bibitem{AntOptimization}
D. Liu, H. Zhang, W. Zheng, and X. Wen, ``The sub-channel allocation algorithm in femtocell networks based on Ant Colony Optimization,'' in {Proc. IEEE MILCOM 2012}, Orlando, FL, pp. 1-6, Oct. 2012.

\bibitem{PathLossModel}
Y. Zhu, Y. Niu, J. Li, D. O. Wu, Y. Li, and D. Jin, ``QoS-aware scheduling for small cell millimeter wave mesh backhaul,'' in {\em Proc. IEEE ICC 2016}, Kuala Lumpur, Malaysia, pp. 1-6, May 2016.

\bibitem{mmwaveblockage}
M. Gao et al., ``Efficient hybrid beamforming with anti-blockage design for high-speed railway communications," {\em IEEE Transactions on Vehicular Technology}, vol. 69, no. 9, pp. 9643-9655, Sept. 2020.

\bibitem{turnoff}
S. Kutty and D. Sen, ``Beamforming for millimeter wave communications: An inclusive survey," {\em IEEE Communications Surveys and Tutorials}, vol. 18, no. 2, pp. 949-973, Secondquarter 2016.

\bibitem{blockconstant}
G. Yang, J. Du and M. Xiao, ``Maximum throughput path selection with random blockage for indoor 60 GHz relay networks," {\em IEEE Transactions on Communications}, vol. 63, no. 10, pp. 3511-3524, Oct. 2015.

\bibitem{beamalign}
V. Va, X. Zhang and R. W. Heath, ``Beam Switching for Millimeter Wave Communication to Support High Speed Trains," 2015 IEEE 82nd Vehicular Technology Conference (VTC2015-Fall), Boston, MA, USA, pp. 1-5, 2015.

\bibitem{Liyan}
L. Yan, X. Fang, L. Hao and Y. Fang, ``A Fast Beam Alignment Scheme for Dual-Band HSR Wireless Networks," in IEEE Transactions on Vehicular Technology, vol. 69, no. 4, pp. 3968-3979, April 2020.

\bibitem{CS}
Y. Li, J. G. Andrews, F. Baccelli, T. D. Novlan and C. J. Zhang, ``Design and Analysis of Initial Access in Millimeter Wave Cellular Networks," in IEEE Transactions on Wireless Communications, vol. 16, no. 10, pp. 6409-6425, Oct. 2017.

\bibitem{RA}
3GPP TR 38.804 V14.0.0, Technical specification group radio access network; Study on new radio access technology; Radio interface protocol aspects (Release 14), Mar. 2017.

\bibitem{preamble}
E. Dahlman, S. Parkvall, and J. Sk¡§old, LTE/LTE-Advanced for Mobile Broadband. Elsevier Ltd, 2011.

\bibitem{WangICDCS2017}
Y. Wang, S. Mao, and T.S. Rappaport, ``On directional neighbor discovery in mmWave networks,'' in {\em Proc. IEEE ICDCS 2017}, Atlanta, GA, pp.1704--1713, June 2017.

\bibitem{HeAccess2015}
Z. He, S. Mao, and T.S. Rappaport, ``On link scheduling under blockage and interference in 60 GHz ad hoc networks,'' {\em IEEE Access Journal}, vol.3, pp.1437-1449, Sept. 2015.

\bibitem{HSRcha1}
J. Yang, B. Ai, D. He, L.Wang, Z. Zhong, and A. Hrovat, ``A simplified multipath component modeling approach for high-speed train channel based on ray tracing," Wireless Commun. Mobile Comput., vol. 2017, no. 333, pp. 1-14, Oct. 2017.

\bibitem{HSRcha2}
D. He et al., ``Channel measurement, simulation, and analysis for high-speed railway communications in 5G millimeter-wave band," {\em IEEE Transactions on Intelligent Transportation Systems}, vol. 19, no. 10, pp. 3144-3158, Oct. 2018.

\bibitem{90GHSRPS}
T. Kawanishi et al., ``Proposal of a new working document of a draft new apt report on millimeter-wave band railway radiocommunication systems between train and trackside and its work plan," in Proc. 20th Meeting
APT Wireless Group, pp. 1-12, 2016.

\bibitem{HSTdelay}
M. Gao et al., ``Edge caching and content delivery with minimized delay for both high-speed train and local users," 2019 IEEE Global Communications Conference (GLOBECOM), Waikoloa, HI, USA, pp. 1-6, 2019.


\bibitem{Duarte}
M.~Duarte, A.~Sabharwal, V.~Aggarwal, R.~Jana, K.~K. Ramakrishnan, C.~Rice, and N.~K. Shankaranayanan, ``Design and characterization of a full-duplex multi-antenna system for WiFi networks,'' \emph{IEEE Transactions on Vehicular Technology}, vol.~63, no.~3, pp. 1160--1177, Mar. 2014.


\bibitem{Jiang17}
Z. Jiang and S. Mao, ``Energy delay trade-off in multi-channel full-duplex wireless LANs,'' {\em IEEE Internet of Things Jounal}, vol.4, no.3, pp.658--669, June 2017.

\bibitem{Wang17}
Y. Wang and S. Mao, ``On distributed power control in full duplex wireless networks,'' {\em Elsevier Digital Communications and Networks Journal}, vol.3, no.1 pp.1--10, Feb. 2017.

\bibitem{Feng15}
M. Feng, S. Mao, and T. Jiang, ``Joint duplex mode selection, channel allocation, and power control for full-duplex cognitive femtocell networks,'' {\em Elsevier Digital Communications and Networks Journal}, vol.1, no.1, pp.30--44, Feb. 2015.


\bibitem{Zhenyu}
Z.~Y. Xiao, P.~F. Xia, and X.~G. Xia, ``Full-duplex millimeter-wave communication,'' \emph{IEEE Wireless Communications Magazine}, vol.~24, no.~6, pp. 136--143, Dec. 2017.

\bibitem{PRINCEANOKYE}
P.~Anokye, R.~K. Ahiadormey, C.~Song, and K.~J. Lee, ``Achievable sum-rate analysis of massive MIMO full-duplex wireless backhaul links in heterogeneous cellular networks,'' \emph{IEEE Access J.}, vol.~6, pp. 23456--23469, May. 2018.

\bibitem{Palomares1976}
U.~M. Garcia Palomares and O.~L. Mangasarian,
``Superlinearly convergent quasi-newton algorithms for nonlinearly constrained optimization problems,''
{\em Mathematical Programming}, vol.11, pp.1--13, Dec. 1976.

\bibitem{ConOpt}
S. Boyd and L. Vandenberghe, Convex Optimization. Cambridge, U.K.: Cambridge Univ. Press, 2004.

\end{thebibliography}
\end{document}